\documentclass[aps,prx,twocolumn,english,groupedaddress, floatfix, longbibliography]{revtex4-2}
\usepackage[utf8]{inputenc}
\usepackage{soul,xcolor}

\usepackage{xcolor}
\usepackage[thicklines]{cancel}
\usepackage{amsmath}
\usepackage{calc}
\usepackage{amsfonts}
\usepackage{graphicx}
\usepackage{dcolumn}
\usepackage{bm}
\graphicspath{{./}}
\newcommand*{\MyDef}{\mathrm{def}}
\newcommand*{\eqdef}{\ensuremath{\mathop{\overset{\MyDef}{=}}}}

\newcommand{\smallsim}{\smallsym{\mathrel}{\sim}}

\makeatletter
\newcommand{\smallsym}[2]{#1{\mathpalette\make@small@sym{#2}}}
\newcommand{\make@small@sym}[2]{%
	\vcenter{\hbox{$\m@th\downgrade@style#1#2$}}%
}
\newcommand{\downgrade@style}[1]{%
	\ifx#1\displaystyle\scriptstyle\else
	\ifx#1\textstyle\scriptstyle\else
	\scriptscriptstyle
	\fi\fi
}
\makeatother
 
\begin{document}
	\setstcolor{red}
	\title{Hamiltonian extrema of an arbitrary flux-biased Josephson circuit}
	\author{A. Miano}
	\email{sandro.miano@yale.edu}
	\affiliation{Department of Applied Physics, Yale University, New Haven, Connecticut 06520, USA}
	\author{V. R. Joshi}
	\affiliation{Department of Applied Physics, Yale University, New Haven, Connecticut 06520, USA}
	\author{G. Liu}
	\affiliation{Department of Applied Physics, Yale University, New Haven, Connecticut 06520, USA}
	\author{W. Dai}
	\affiliation{Department of Applied Physics, Yale University, New Haven, Connecticut 06520, USA}
	\author{P. D. Parakh}
	\affiliation{Department of Applied Physics, Yale University, New Haven, Connecticut 06520, USA}
	\author{L. Frunzio}
	\affiliation{Department of Applied Physics, Yale University, New Haven, Connecticut 06520, USA}
	\author{M. H. Devoret}
	\email{michel.devoret@yale.edu}
	\affiliation{Department of Applied Physics, Yale University, New Haven, Connecticut 06520, USA}
	\begin{abstract}
		Flux-biased loops including one or more Josephson junctions are ubiquitous elements in quantum information experiments based on superconducting circuits. These quantum circuits can be tuned to implement a variety of Hamiltonians, with applications ranging from decoherence-protected qubits to quantum limited converters and amplifiers. The extrema of the Hamiltonian of these circuits are of special interest because they govern their low-energy dynamics. However, the theory of superconducting quantum circuits lacks so far a systematic method to find these extrema and compute the series expansion of the Hamiltonian  in their vicinity for an arbitrary non-linear superconducting circuit. We present such method, which can aid the synthesis of new functionalities in quantum devices.
	\end{abstract}
	\maketitle
	\section{Introduction}
\begin{figure}
		\centering
		\includegraphics[width = \columnwidth]{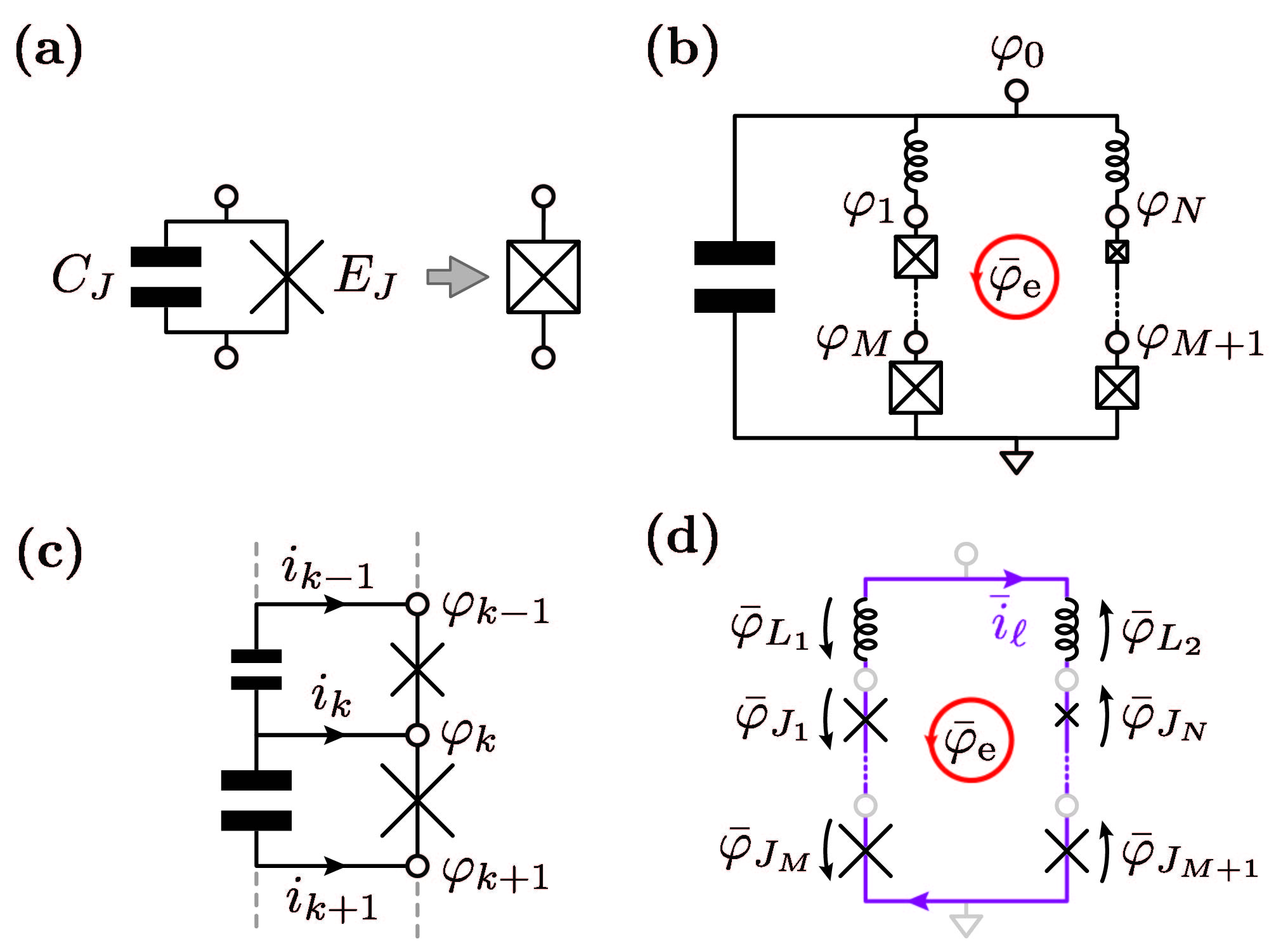}
		\caption{Electrical properties of Josephson flux-biased superconducting dipoles.
			(a) A JJ is characterized by a tunneling element (cross symbol) of energy $E_J$, shunted by a capacitance $C_J$ modeling the tunnel barrier. The combination of the two elements is represented as a cross-in-box symbol.
			(b) Arbitrary flux-biased superconducting dipole shunted by an external capacitor.
			It consists of linear geometric inductances and JJs, threaded by an external DC magnetic phase $\bar{\varphi}_\mathrm{e}$. Including the JJs' capacitances, this circuit has $N+1$ active nodes, and is thus described by a $N+1$-body Hamiltonian where the main node is described by the phase variable $\varphi_0$ and the $k$-th internal node is described by the phase variable $\varphi_k$.
			(c) The $k$-th component of the vector $\nabla_{\mathbf{\varphi}}U_\mathrm{ind}$ is proportional to the total current entering the inductive subnetwork at node $k$.
			(d) At equilibrium, the loop in (b) is threaded by a purely DC loop current $i_\mathrm{\ell}$, as the currents flowing through the capacitances are suppressed.
			Consequently, the $x$-th JJ in the loop has an equilibrium phase drop $\bar{\varphi}_{J_x}$ and the linear inductors in left and right branches have, respectively, equilibrium phase drops $\bar{\varphi}_{L_1}$ and $\bar{\varphi}_{L_2}$.
			In this figure, the reference direction for the phase drops is taken to align with the reference direction for the loop current.}
		\label{fig:1}
	\end{figure}

	State-of-the-art superconducting quantum processors rely on nonlinear circuits operating at GHz frequencies. These devices typically consist of a combination of linear microwave circuitry and Josephson tunnel junctions (JJs) \cite{vool_introduction_nodate}. 
	Specifically, a JJ behaves as a nonlinear inductance described by a potential energy function $U(\varphi) = -E_J\cos{\varphi}$, where $E_J$ is the Josephson energy and $\varphi$ is the phase drop across the JJ, itself linearly related to the integral of the voltage across the element \cite{noauthor_high_1982}. The JJ is also characterized by a capacitance $C_J$ in parallel with the Josephson element, modeling the influence of  the tunnel barrier. This defines a charging energy $E_C = e^2/(2C_J)$, where $e$ is the electron charge.
	Such a model is rendered in Fig. \ref{fig:1}(a), where a JJ is represented as a cross-in-box symbol. The cross symbol is associated with the nonlinear inductive component of a JJ, while the box represents its capacitive shunt.
	One or multiple JJs, together with superconducting wires, can be combined to form a superconducting loop which can be threaded by an external DC magnetic flux $\bar{\Phi}_\mathrm{e}$. To ease the notation, in the rest of the manuscript we will use the normalized external DC magnetic flux, defined as 
	\begin{equation}
		\bar{\varphi}_\mathrm{e}\eqdef 2\pi \frac{\bar{\Phi}_\mathrm{e}}{\Phi_0},
	\end{equation}
	where $\Phi_0 = h/2e$ is the magnetic flux quantum.
	This flux acts as a tuning parameter which controls the linear and nonlinear properties of the loop. 
	By properly assembling such loops and choosing a particular value of $\bar{\varphi}_\mathrm{e}$, it is possible to implement a large variety of quantum systems, ranging from noise-protected qubits \cite{manucharyan_fluxonium_2009,smith_superconducting_2020,gyenis_experimental_2021,kitaev_protected_2006, grimm_stabilization_2020, frattini_squeezed_2022} to readout circuits as quantum-limited amplifiers \cite{castellanos-beltran_amplification_2008, ranadive_kerr_2022, zorin_traveling-wave_2017,abdo_nondegenerate_2013, abdo_directional_2013, frattini_3-wave_2017, sivak_josephson_2020} and control circuits such as parametric couplers \cite{zhou_modular_2022, chapman_high_2022}.
	Superconducting loops are usually modeled as $N$-body quantum mechanical systems, where $N$ is the number of JJs in the loop.
	The $k$-th JJ is characterized by an inductive energy $E_{J_k}$ and a charging energy $E_{C_k}$.
	In addition, each loop hosts a linear inductor representing its total geometrical inductance.
	As a consequence, any superconducting loop could host up to $N$ resonant modes.
	Typically, a superconducting loop is coupled to external circuitry via two of its terminals \cite{castellanos-beltran_amplification_2008, frattini_3-wave_2017}. 
	For instance, a loop can be shunted by an external capacitor of charging energy $E_C$, as in Fig. \ref{fig:1}(b). This defines a reference (ground) node and a main active node (top node) for the loop, as well as two branches. Each branch corresponds to a path from the main active node to the reference node.
	In the rest of the manuscript, we will refer to the two-terminal device obtained by selecting two nodes of the loop as a \emph{dipole}.
	Note that a loop hosting $N$ elements can generate up to $N-1$ different dipoles.
	A general expression for the classical Hamiltonian of the capacitively shunted loop in Fig. \ref{fig:1}(b) is given by a combination of the capacitive (kinetic) energy $T_\mathrm{cap}$ and inductive (potential) energy $U_\mathrm{ind}$
	\begin{equation}
		\label{eq:generic_hamiltonian}
		H(\boldsymbol{n}, \boldsymbol{\varphi}, \bar{\varphi}_\mathrm{e}) = T_\mathrm{cap}(\boldsymbol{n}) + U_\mathrm{ind}(\boldsymbol{\varphi}, \bar{\varphi}_\mathrm{e})
	\end{equation}
	where $\boldsymbol{\varphi} = [\varphi_0,\varphi_1,\dots, \varphi_N]$ and $\boldsymbol{n} = [n_0,n_1,\dots, n_N]$ are the vectors representing the phase and conjugate charge variables associated with the active nodes of the loop, respectively \cite{vool_introduction_nodate}. Note that the $\varphi_k$  are linearly related to the integral of the voltage between node $k$ and the ground node. The phase $\varphi_0$ corresponds to the main active node of the dipole, while the others correspond to the internal nodes of the dipole.
	The Hamiltonian \eqref{eq:generic_hamiltonian} can then be quantized by promoting the variables $\varphi_k$ and $n_k$ to the operators $\hat{\varphi}_k$ and $\hat{n}_k$, respectively, satisfying the commutation relation $[\hat{\varphi}_k,\hat{n}_k] = i$. The resulting Hamiltonian operator $\hat{H}$ is related to the classical Hamiltonian \eqref{eq:generic_hamiltonian} via the relation
	\begin{equation}
		\label{eq:generic_hamiltonian_quantized}
		\hat{H} = H(\hat{\boldsymbol{n}}, \hat{\boldsymbol{\varphi}}, \bar{\varphi}_\mathrm{e})
	\end{equation}
	where $\hat{\boldsymbol{\varphi}} = [\hat{\varphi}_0,\hat{\varphi}_1,\dots, \hat{\varphi}_N]$ and $\hat{\boldsymbol{n}} = [\hat{n}_0,\hat{n}_1,\dots, \hat{n}_N]$ are the vectors of phase and charge operators, respectively.
	For a given external flux-bias $\bar{\varphi}_\mathrm{e}$, the eigenstates of $\hat{H}$ can be computed and further analyzed to yield the properties of the system. This comprehensive approach can be extended to an arbitrary superconducting circuit, and is currently at the core of many general-purpose quantum circuits simulators \cite{groszkowski_scqubits_2021, gely_qucat_2020, aumann_circuitq_2022}. 
	However, this description does not directly capture the number and positions of the local extrema of the classical potential energy $U_\mathrm{ind}(\boldsymbol{\varphi}, \bar{\varphi}_\mathrm{e})$ in Eq. \eqref{eq:generic_hamiltonian}, the knowledge of which is very convenient when designing a device for a particular purpose. 
	For instance, many topological protected qubits \cite{smith_superconducting_2020,gyenis_experimental_2021,kitaev_protected_2006} are implemented with Hamiltonians whose ground state wavefunction, in the phase representation, can be imagined as spread across multiple minima of $U_\mathrm{ind}$. 
	On the other hand, quantum-limited amplifiers and couplers typically require the presence of a single  minimum for $U_\mathrm{ind}$ to reliably implement the multi-photon parametric processes at their core.
	These devices are typically described by a series-expansion of their effective Hamiltonian around the minimum.
	A brute-force approach to determining the number and positions of the potential energy local extrema is to compute, for a fixed $\bar{\varphi}_\mathrm{e}$, the \emph{equilibrium points} of the circuit \cite{minev_energy-participation_2021}, defined as the phase vector $\bar{\boldsymbol{\varphi}} =  [\bar{\varphi}_0,\bar{\varphi}_1,\dots, \bar{\varphi}_N]$ which satisfies the set of equations 
	\begin{equation}
		\label{eq:equilibrium_points}
		\left.\nabla_\mathrm{\boldsymbol{\varphi}}U_\mathrm{ind}\right|_{\bar{\boldsymbol{\varphi}}} = \boldsymbol{0}
	\end{equation}
	where $\nabla_\mathrm{\boldsymbol{\varphi}}$ is the gradient with respect to the components of $\boldsymbol{\varphi}$.
	In the general case, this method requires solving a set of transcendental trigonometric equations with iterative root-finding algorithms.
	A search for all the extrema of a potential energy function is thus very challenging and time consuming.
	
	Here, we propose a systematic method to describe the equilibrium properties of an arbitrary flux-biased superconducting dipole.
	The essence of our method is based on the relation
	\begin{equation}
		\label{eq:nabla_u_to_current}
		\frac{\partial U_\mathrm{ind}}{\partial\varphi_k} = \frac{\Phi_0}{2\pi}i_k(\boldsymbol{\varphi}),
	\end{equation}
	where $i_k(\boldsymbol{\varphi})$ is the total current flowing into the inductive subnetwork through the $k\mathrm{-th}$ active node, as shown in Fig. \ref{fig:1}(c).
	This last relation can be demonstrated by computing the instantaneous power entering the inductive subnetwork as
		\begin{equation}
			\label{eq:total_inductive_power}
			P_\mathrm{ind} = \frac{dU_\mathrm{ind}}{dt} = \sum_{k=0}^{N}i_k(\boldsymbol{\varphi})V_k
		\end{equation}
	where $V_k$ is the voltage between node $k$ and the ground node. Being
	\begin{equation}
		\label{eq:voltage_phase}
		V_k = \frac{\Phi_0}{2\pi}\frac{d\varphi_k}{dt},
	\end{equation}
	the total differential of $U_\mathrm{ind}$ can be computed from Eq. \eqref{eq:total_inductive_power} as
	\begin{equation}
		\label{eq:potential_total_differential}
		dU_\mathrm{ind} = \frac{\Phi_0}{2\pi}\sum_{k=0}^N i_k(\boldsymbol{\varphi})d\varphi_k
	\end{equation}
	from which expression \eqref{eq:nabla_u_to_current} arises.
	Consequently, Eq. \eqref{eq:equilibrium_points} imposes that, at each equilibrium point, $i_k(\boldsymbol{\bar{\varphi}})=0$, thus only a persistent loop current \cite{smith_observation_1965} $\bar{i}_\ell$ is allowed to circulate in the dipole, as represented in Fig. \ref{fig:1}(d).
	As a consequence, the $x$-th element in the loop will have an equilibrium phase drop $\bar{\varphi}_x$, such that $i_x(\bar{\varphi}_x)=\bar{i}_\mathrm{\ell}$ \cite{minev_energy-participation_2021}, where $i_x(\varphi_x)$ is the current-phase relation (CPR) describing the $x$-th element, where $x\in\{{L_1,L_2,J_1,\dots,J_N}\}$.
	In particular, the CPR of a JJ reads
	\begin{equation}
		i(\varphi) = I_C\sin{\varphi}
	\end{equation}
	while that of a linear inductor reads
		\begin{equation}
		i(\varphi) = \frac{\Phi_0}{2\pi L}\varphi
	\end{equation}
	where $I_C$ is the critical current of the JJ and $L$ is the inductance of the linear inductor.
	We want to emphasize that all the electrical variables at equilibrium are time-independent: as a consequence, their properties are independent from the capacitive subnetwork of the dipole in Fig. \ref{fig:1}(b).
	
	In section II, we show how the equilibrium points of a single loop are univocally related to the CPR of a single, open branch constructed from a sequence of all the elements in the loop. 
	Then, we show how a generic CPR can be expressed as a parametric relation. 
	To obtain such a representation, the curvilinear parameter of choice is the phase drop across the JJ with the smallest critical current, which we name the \emph{free JJ} for reasons that will become clear later.
	In section III, we study the multistability of a superconducting loop and derive a set of rules that quantify the properties of circuits with multiple equilibrium points.
	In section IV, we discuss the application of our method to describe the low energy properties of the quantum Hamiltonian \eqref{eq:generic_hamiltonian_quantized} which, when $E_C\ll{E_{C_k}}$, can be assumed independent from the JJs' capacitances. 
	Under such approximation, we introduce an effective potential energy function associated to $U_\mathrm{ind}$, whose series expansion coefficients can be computed by systematically combining those of the inductive elements forming the dipole.
	Finally, in section V we discuss the applications of our approach: optimization and synthesis of superconducting circuits, modeling the effect of fabrication uncertainties and tight-binding approximation for multi-body superconducting circuits.
	
	We have implemented the results of this work in a Python package available on a GitHub repository \cite{nina}, named `Nonlinear Inductive Network Analyzer' (NINA).
	The repository includes examples for common superconducting dipoles, as well as a comparison between our method and the method used in previous literature to analyze a SNAIL \cite{frattini_3-wave_2017}.
	\section{Equilibrium points of an arbitrary superconducting loop}
	
\begin{figure}
		\centering
		\includegraphics[width = \columnwidth]{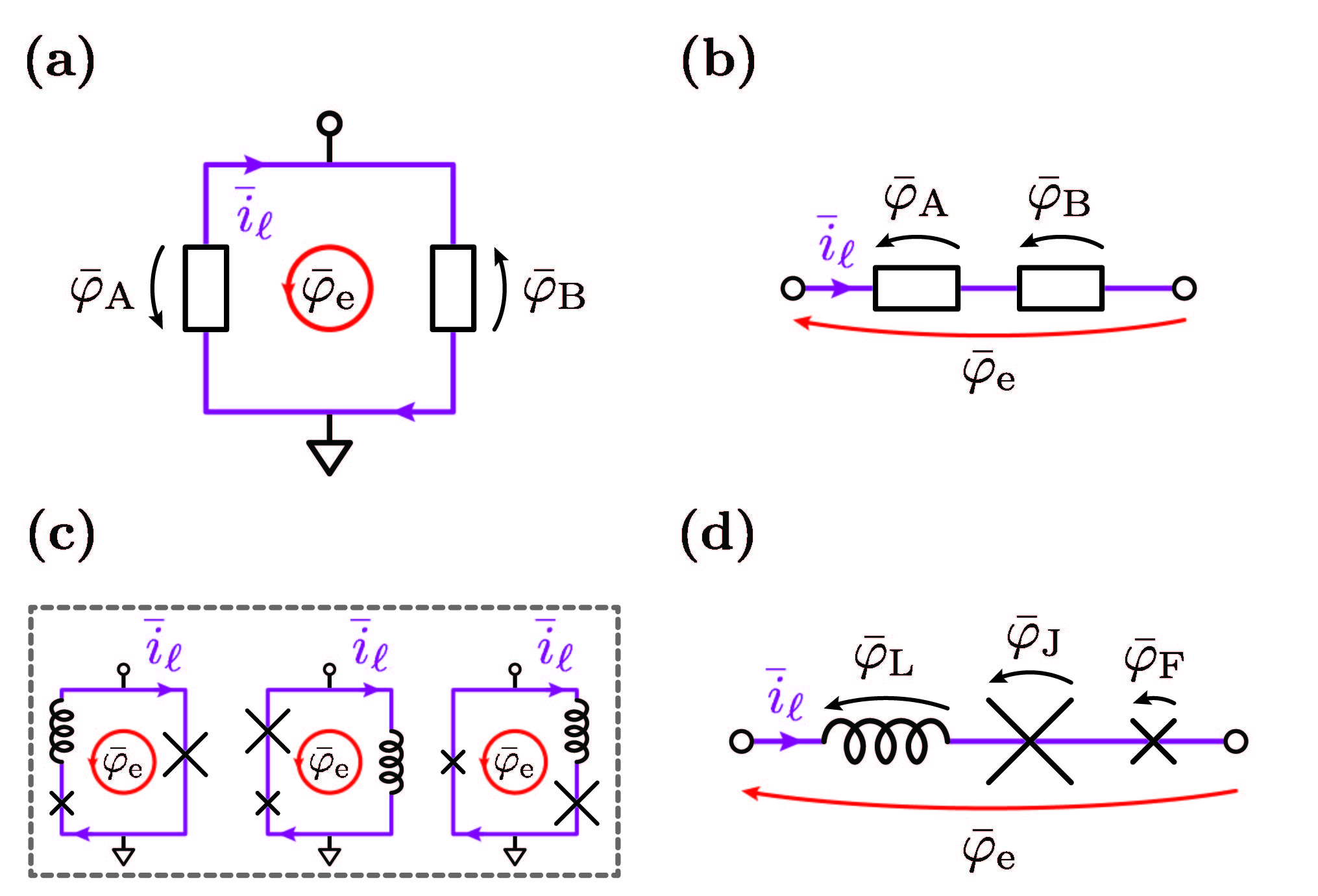}
		\caption{Steady state representation of a superconducting loop. 
			(a)
			A superconducting loop formed by two branches A and B is threaded by an external DC magnetic flux $\bar{\varphi}_\mathrm{e}$. At equilibrium, the left and right branches have, respectively, phase drops $\bar{\varphi}_\mathrm{A}$ and $\bar{\varphi}_\mathrm{B}$.
			(b) 
			Equivalent open branch associated with the loop in panel (a). The equations describing the equilibrium solutions for the loop in panel (a) are the same as for the single open branch in (b) consisting of \textbf{A} and \textbf{B} in series, provided that the net current flowing through the open branch corresponds to the loop current $\bar{i}_\ell$ and the total phase drop across the open branch corresponds to the external flux $\bar{\varphi}_\mathrm{e}$.
			(c)
			Three dipoles obtained by permuting three different elements share the same equilibrium points. They all map to the same equivalent open branch. 
			(d)
			The equivalent open branch associated with each dipole in (c) is the series of all the elements in the loop.}
		\label{fig:2}
	\end{figure}

	As mentioned in section I, the equilibrium point of a superconducting loop threaded by an external DC flux-bias $\bar{\varphi}_\mathrm{e}$ is characterized by the flow of a purely persistent current $\bar{i}_\ell$.
	For the arbitrary superconducting loop in Fig. \ref{fig:2}(a), the Kirchhoff laws read
	\begin{equation}
		\label{eq:single_loop_KL}
		\left\{
		\begin{aligned}
			&i_\mathrm{A}(\bar{\varphi}_\mathrm{A}) = i_\mathrm{B}(\bar{\varphi}_\mathrm{B}) = \bar{i}_\ell\\
			&\bar{\varphi}_\mathrm{A} + \bar{\varphi}_\mathrm{B} = \bar{\varphi}_\mathrm{e}.
		\end{aligned}
		\right.
	\end{equation}
	Remarkably, the set of equations \eqref{eq:single_loop_KL} does also describe the case where the two branches \textbf{A} and \textbf{B} are arranged in series, as depicted in Fig. \ref{fig:2}(b). We call such configuration the \emph{equivalent open branch} of the loop. 
	From the point of view of the persistent current $\bar{i}_\ell$, the two branches \textbf{A} and \textbf{B} can be regarded to be in series. 
	This observation has some important consequences. For instance, as represented in Fig. \ref{fig:2}(c), a set of three different inductive elements can be arranged to form three different flux-biased dipoles.
	These dipoles are obtained by selecting different pairs of terminals from the same loop. Consequently, the equilibrium phases of each element will be the same for a common value of $\bar{\varphi}_\mathrm{e}$, regardless of the terminals arrangement.
	The three superconducting dipoles would however differ by the shape of their potential energy function, as explained in section IV.
	
\begin{figure}
		\centering
		\includegraphics[width = \columnwidth]{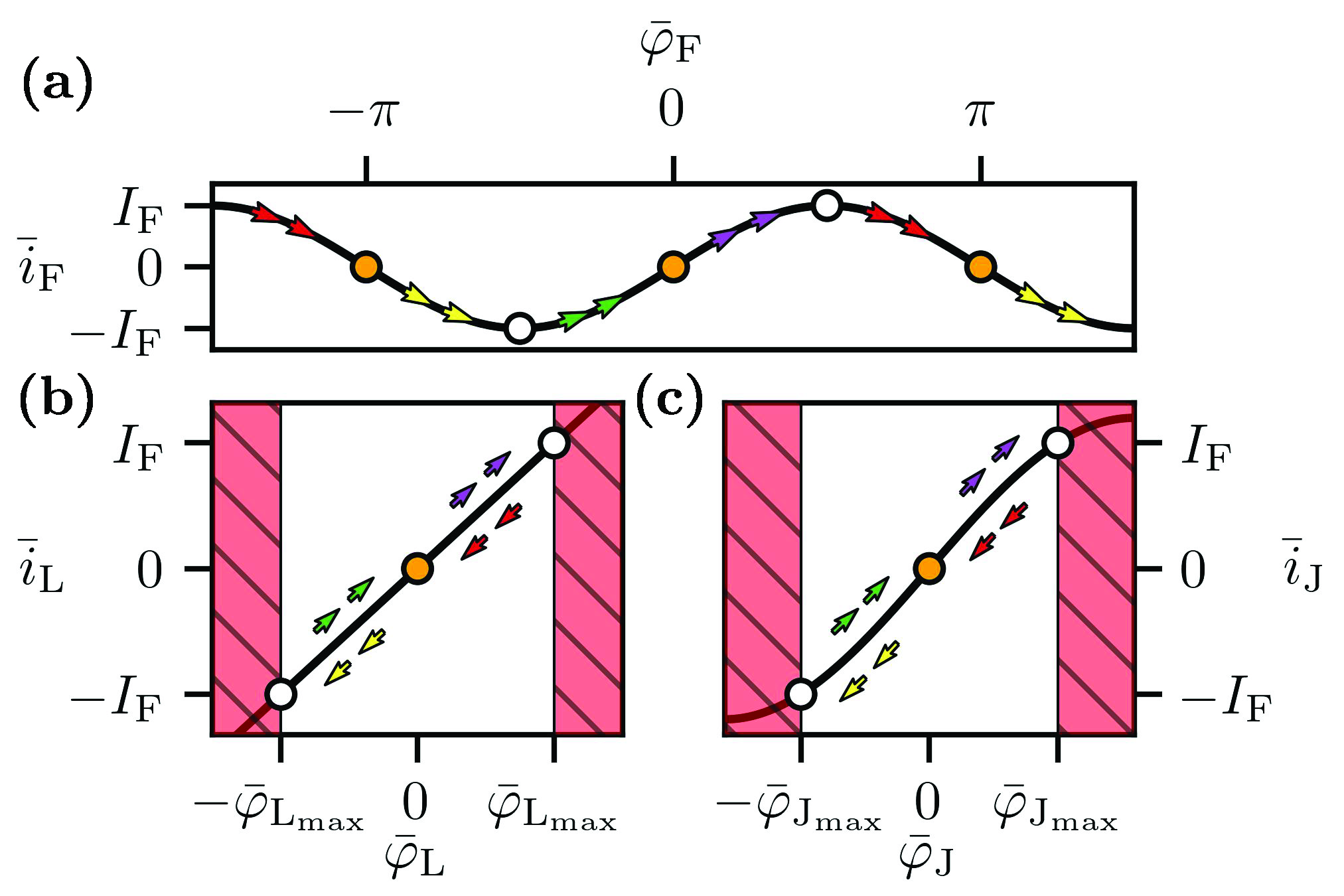}
		\caption{Current-phase relations for the elements of the circuit in the right panel of Fig. \ref{fig:2}(c).
			(a) Current-phase relation for the \emph{free JJ}. Being the current-limiting element, the domain of variation of its phase is unconstrained, hence the term ``free".
			(b) Current-phase relation for the linear inductor. The presence of a free JJ in series constraints the inductor phase drop in the interval $[-\bar{\varphi}_\mathrm{L_{max}}, \bar{\varphi}_\mathrm{L_{max}}]$.
			(c) Current-phase relation for the largest JJ. Similarly, the free JJ in series constrains the phase drop in the interval $[-\bar{\varphi}_\mathrm{J_{max}}, \bar{\varphi}_\mathrm{J_{max}}]$.}
		\label{fig:3}
	\end{figure}

	To analyze the equivalent open branch in Fig. \ref{fig:2}(d), we introduce the maximum DC current allowed through such branch, $I_F$, corresponding to the critical current of the free JJ.
	While the phase of the free JJ is unbounded, as rendered in Fig. \ref{fig:3}(a), the domain of variation of the phase across the other elements is constrained.
	The absolute value of the maximum phase drop across the linear inductor and the larger JJ will be given, respectively, by $\textrm{max}\{|\bar{\varphi}_L|\} = \beta_L$ and $\textrm{max}\{|\bar{\varphi}_J|\} = \arcsin{\beta_J}$, where 
	\begin{equation}
		\begin{aligned}
			\beta_L & = \frac{2\pi LI_F}{\Phi_0}\\
			\beta_J & = \frac{I_F}{I_J}.
		\end{aligned}
	\end{equation}
	Note that $\beta_J \leq 1$ by definition.
	In these last definitions, $L$ is the inductance of the linear inductor, and $I_J$ the critical current of the big JJ.
	Such maximum values correspond, as in Fig. \ref{fig:3}(b) and (c), to a flow of current through the elements equal to $I_F$, i.e. to a phase drop across the small JJ, $\bar{\varphi}_F = \pi/2$.
	For a generic value of $\bar{\varphi}_F$, the phases of the constrained elements read
	\begin{equation}
		\label{eq:phases_constrained}
		\begin{aligned}
			&\bar{\varphi}_L = \beta_L\sin{\bar{\varphi}_F}\\
			&\bar{\varphi}_J = z\pi + (-1)^z\arcsin{\left(\beta_J\sin{\bar{\varphi}_F}\right)},
		\end{aligned}
	\end{equation}
	where $z\in\mathbb{Z}$ is the integer multiple of $\pi$ around which the CPR of the bigger JJ is evaluated. 
	In the rest of the manuscript, we will only consider the case $z=0$, as in Fig. \ref{fig:3}(c). 
	The properties of a superconducting array for $z\neq0$ are outside the scope of this work, since they are associated to higher potential energy curves. These curves can still have equilibrium points with positive curvature, but of higher energy with respect to those of the potential energy associated with $z=0$. As a consequence, the equilibrium points for $z\neq0$ can be classified as metastable.
	
	We can now establish the CPR of the branch, observing that the total phase drop $\bar{\varphi}$ is obtained by adding the phase drops across the elements, while the current $\bar{i}$ can be expressed as the one through the free JJ. 
	We obtain the following system of parametric equations
	\begin{equation}
		\label{eq:CPR_parametric}
		\left\{
		\begin{aligned}
			&\bar{\varphi} = \bar{\varphi}_F + \beta_L\sin{\bar{\varphi}_F} + \arcsin{\left(\beta_J\sin{\bar{\varphi}_F}\right)}\\
			&\bar{i} = I_F\sin{\bar{\varphi}_F}
		\end{aligned}
		\right.
	\end{equation}
	\begin{figure}
		\centering
		\includegraphics[width = \columnwidth]{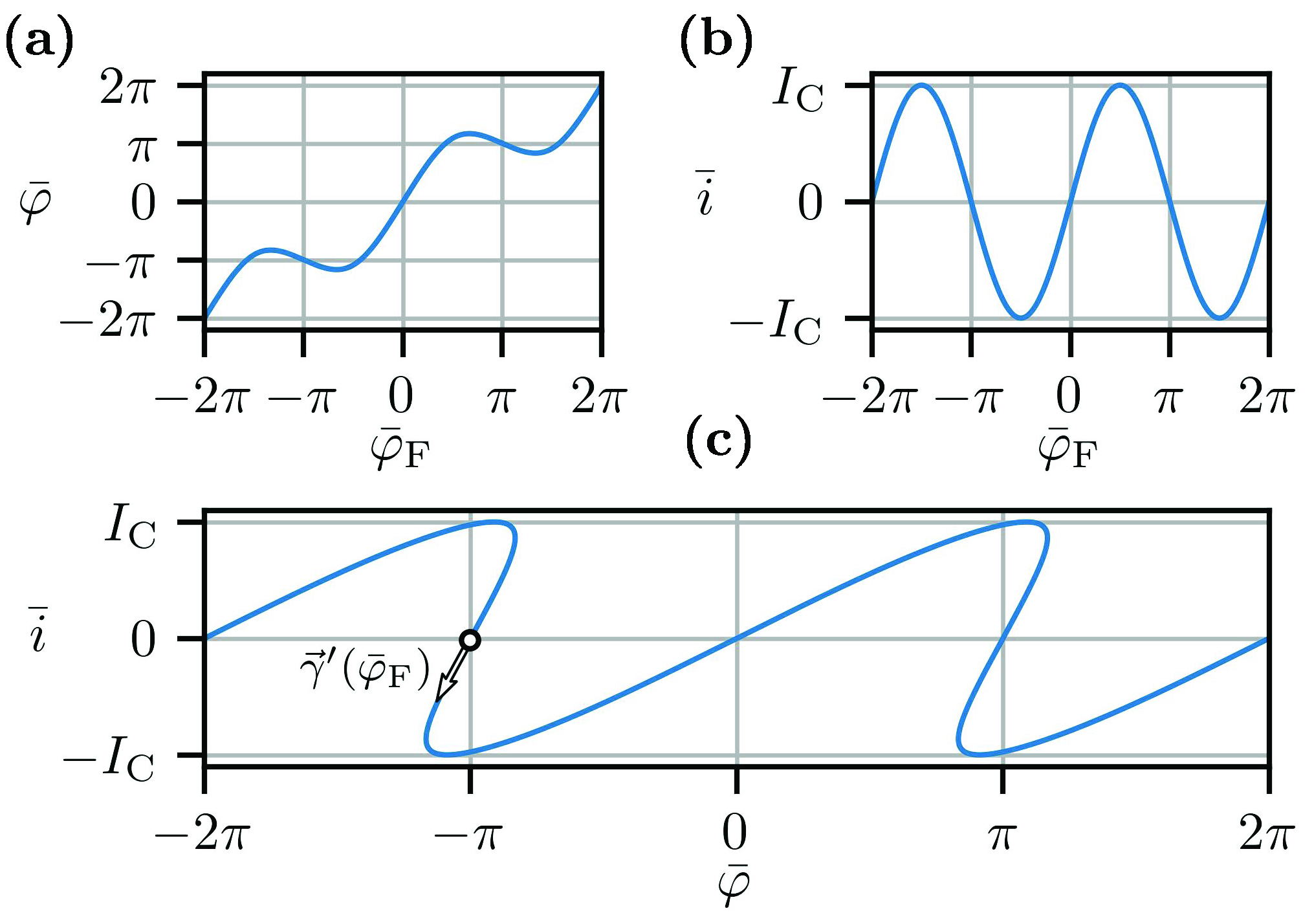}
		\caption{Current-phase relation of the branch in Fig. 2(d) represented as a parametric relation.
			(a) Total phase drop across the branch as a function of the free JJ equilibrium phase.
			(b) Current through the branch as a function of the free JJ equilibrium phase.
			(c) CPR of the branch obtained as a parametric curve $\vec{\gamma}(\varphi_F)$ of curvilinear parameter $\varphi_F$. 
			The CPR is a multi-valued function if its tangent vector $\vec{\gamma}^\prime(\bar{\varphi}_F)$ has a negative horizontal component for $\bar{\varphi}_F=\pi$.}
		\label{fig:4}
	\end{figure}
	which represents the implicit form of the CPR of the equivalent open branch in Fig. \ref{fig:2}(d) as a 1D parametric curve $\vec{\gamma}(\bar{\varphi}_F) = \langle\varphi(\bar{\varphi}_F),i(\bar{\varphi}_F)\rangle$, with the curvilinear parameter being $\bar{\varphi}_F$. 
	Note that the total phase drop $\bar{\varphi}$ across an equivalent open branch corresponds to the external flux of the associated loop.
	In Fig. \ref{fig:4}, the $\bar{\varphi}(\bar{\varphi}_F)$, $\bar{i}(\bar{\varphi}_F)$ and $\bar{i}(\bar{\varphi})$ are computed for $\beta_L = 1$ and $\beta_J = 0.25$. 
	
	The description discussed so far can be extended to the case where two JJs are identical, i.e. $\beta_J = 1$. 
	In such case, only one of the two JJs should be elected to be the free one, while the other has to be treated as constrained.
	
	To generalize the system \eqref{eq:CPR_parametric} to the case of $N$ larger JJs, it is necessary to include all their contributions to the total phase $\bar{\varphi}$. The simple implicit CPR \eqref{eq:CPR_parametric} then becomes the general
	\begin{equation}
		\label{eq:general_CPR}
		\left\{
		\begin{aligned}
			&\bar{\varphi} = \bar{\varphi}_F + \beta_L\sin{\bar{\varphi}_F} + \sum_{k=0}^N\arcsin{\left(\beta_{J_k}\sin{\bar{\varphi}_F}\right)}\\
			&\bar{i} = I_F\sin{\bar{\varphi}_F}
		\end{aligned}
		\right.
	\end{equation}
	where $\beta_{J_k} = I_F/I_{J_k}$.
	Since the phases across the constrained elements are analytical functions of $\bar{\varphi}_F$ as in \eqref{eq:phases_constrained}, the set of equations \eqref{eq:general_CPR} describes the equilibrium properties of a generic superconducting loop for an arbitrary value of external flux when imposing $\bar{\varphi} = \bar{\varphi}_\mathrm{e}$ and $\bar{i} = \bar{i}_\ell$. 
	The only caveat is that the inverse function $\bar{\varphi}_F(\bar{\varphi})$, in general, does not admit an analytical expression. However, it can be approximated via numerical interpolation techniques, up to a desired numerical precision.
	
	This method can also be extended to include nonlinear inductive elements with non-sinusoidal CPRs as nanowires \cite{spanton_currentphase_2017} or JJs with higher Josephson harmonics \cite{willsch2023observation}.
	
	\section{Multi-minima superconducting loops}
	Some loop designs, for instance the fluxonium \cite{manucharyan_fluxonium_2009} and $\cos{2\varphi}$ \cite{smith_superconducting_2020} qubits, are purposely designed to work in a multi-minima configuration, while others require the presence of a single operating point, as in the case of parametric couplers \cite{zhou_modular_2022, chapman_high_2022} and amplifiers \cite{castellanos-beltran_amplification_2008, bergeal_phase-preserving_2010, frattini_3-wave_2017}. To formulate the conditions under which an arbitrary flux-biased superconducting loop has one or multiple operating points, it is useful to acknowledge that its operating points can be determined by analyzing the CPR of its equivalent open branch. 
	In particular, we note how, as displayed in Fig. \ref{fig:4}, $\bar{i}(\bar{\varphi})$ is a single-valued function if the component of $\vec{\gamma}^\prime(\pi)$ along the $\bar{\varphi}_F$ axis is positive, or a multi-valued function if the same component of $\vec{\gamma}^\prime(\pi)$ is negative.
	We want to clarify that, even if Fig. \ref{fig:4} describes a particular associated branch, such observation applies to the general case: the generic CPR is a continuous curve, and $\bar{\varphi}(\bar{\varphi}_F = \pi) = \pi$ regardless of the branch complexity, as can be obtained from the first equation in \eqref{eq:general_CPR}.
	The component of $\vec{\gamma}^\prime(\bar{\varphi}_F)$ along the $\bar{\varphi}_F$ axis is given by
	\begin{equation}
		\label{eq:CPR_directional_derivative}
		\frac{d\bar{\varphi}}{d\bar{\varphi}_F} = 1 + \cos{\bar{\varphi}_F}\left(\beta_L + \sum_{k=0}^N \frac{\beta_{J_k}}{\sqrt{1 - \beta^2_{J_k}\sin^2{\bar{\varphi}_F}}}\right)
	\end{equation}
	which, evaluated for $\bar{\varphi}_F = {\pi}$, results in
	\begin{equation}
		\label{eq:CPR_directional_derivative_at_pi}
		\left.\frac{d\bar{\varphi}}{d\bar{\varphi}_F}\right|_{\bar{\varphi}_F = \pi} = 1 - \beta_\mathrm{tot},
	\end{equation}
	where we have defined 
	\begin{equation}
		\label{eq:beta_tot}
		\beta_\mathrm{tot} = \beta_L + \sum_{k=0}^N \beta_{J_k}.
	\end{equation}
	Consequently, the conditions for single- or multi- valued branches are
	\begin{equation}
		\label{eq:multistability_criterion}
		\begin{aligned}
			&\beta_\mathrm{tot} \leq 1 \rightarrow \mathrm{single\;minimum}\\
			&\beta_\mathrm{tot} > 1 \rightarrow \mathrm{multiple\;minima}
		\end{aligned}
	\end{equation}
	These inequalities generalize those already known for common loops as rf-SQUIDs \cite{zorin_josephson_2016} and SNAILs \cite{frattini_3-wave_2017} to an arbitrary superconducting loop.
	\begin{figure}
		\centering
		\includegraphics[width = \columnwidth]{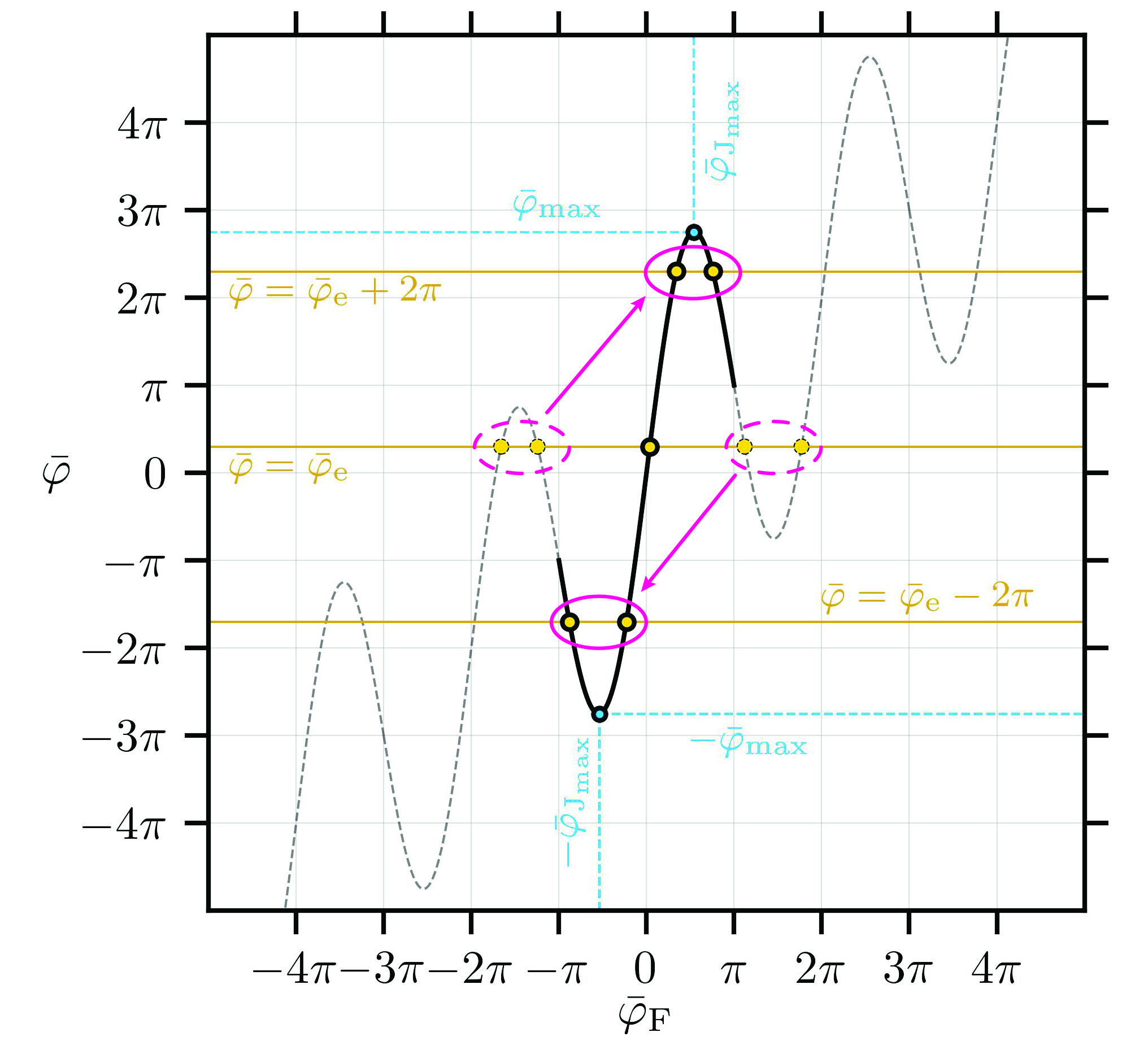}
		\caption{Equilibrium points of the free JJ of a superconducting branch for a given value of total phase drop $\bar{\varphi}$. 
			Interpreting the branch as the \emph{equivalent open branch} of a loop threaded with a DC magnetic flux $\bar{\varphi}_\mathrm{e}$, then $\bar{\varphi} = \bar{\varphi}_\mathrm{e}$.
			When a loop is multistable, the equilibrium points of the free JJ can fall outside the interval $[-\pi,\pi)$ (yellow dots in dashed ovals). However, the translational symmetry of the function $\bar{\varphi}(\bar{\varphi}_F)$ can be exploited to map all the free JJ equilibrium points back in the interval $[-\pi,\pi)$ (yellow dots in solid ovals), by introducing a grid of effective external fluxes $\bar{\varphi}_\mathrm{e} + 2n\pi$. Said $\bar{\varphi}_\mathrm{max}$ the local maxima of the function $\bar{\varphi}(\bar{\varphi}_F)$, it is possible to count and locate all the equilibrium points available in an arbitrary loop for a given value of external flux.}
		\label{fig:5}
	\end{figure}
	In the multi-minima scenario $\beta_\mathrm{tot} > 1$, for a given value of $\bar{\varphi}_\mathrm{e}$ there will be a set of equilibrium free phases, each one corresponding to an equilibrum point of the loop. Formally, such solutions are those of the system
	\begin{equation}
		\label{eq:_multiminima_system}
		\left\{
		\begin{aligned}
			&\bar{\varphi}(\bar{\varphi}_F) = \bar{\varphi}_\mathrm{e}\\
			&\bar{\varphi}_F \in \; (-\infty,\infty)
		\end{aligned}
		\right.
	\end{equation}
	With reference to Fig. \ref{fig:5}, the solutions of this system are the yellow dots at intersection between the $\bar{\varphi}(\bar{\varphi}_F)$ curve and the yellow line $\bar{\varphi} = \bar{\varphi}_\mathrm{e}$.
	To count the number of operating points in an arbitrary loop for any value of external flux, it is useful to note how the function $\bar{\varphi}(\bar{\varphi}_F)$ has the translational symmetry $\bar{\varphi}(\bar{\varphi}_F + 2\pi) = 2\pi + \bar{\varphi}(\bar{\varphi}_F)$, as can be verified from the first equation in \eqref{eq:general_CPR}. Consequently, the solutions of the system \eqref{eq:_multiminima_system} can be mapped to those of an equivalent system
	\begin{equation}
		\label{eq:_multiminima_equivalent_system}
		\left\{
		\begin{aligned}
			&\bar{\varphi}(\bar{\varphi}_F) = \bar{\varphi}_\mathrm{e} + 2n\pi \;\;\;\;\; n\in\mathbb{Z}\\
			&\bar{\varphi}_F \in [-\pi,\pi)
		\end{aligned}
		\right.
	\end{equation}
	which has the benefit of limiting the range of $\bar{\varphi}_F$ where to search for solutions.
	The transition between system \eqref{eq:_multiminima_system} and \eqref{eq:_multiminima_equivalent_system} is also represented in Fig. \ref{fig:5}, where all the solutions of \eqref{eq:_multiminima_system} which fall outside the interval $\bar{\varphi}_F \in (-\infty,-\pi) \cup [\pi,\infty)$ (enclosed in a dashed oval) can be mapped to equivalent solutions in the interval $\bar{\varphi}_F \in [-\pi,\pi)$ (enclosed in a solid oval).
	Moreover,  $\bar{\varphi}(\bar{\varphi}_F)$ being a bounded function for $\bar{\varphi}_F \in [-\pi,\pi)$, only a finite number of $n$ need to be considered to find all the solutions of the equivalent system \eqref{eq:_multiminima_equivalent_system}. 
	With reference to Fig. \ref{fig:5}, the upper- and lower-bound on $n$ for a given value of $\bar{\varphi}_\mathrm{e}$ are respectively given by
	\begin{equation}
		\label{eq:number_of_solutions_0}
		\begin{aligned}
			&N_\uparrow = 
			\left\lfloor\frac{\bar{\varphi}_\mathrm{e} - \bar{\varphi}_\mathrm{max}}{2\pi}\right\rfloor\\
			&N_\downarrow = - \left\lfloor\frac{\bar{\varphi}_\mathrm{e} + \bar{\varphi}_\mathrm{max}}{2\pi}\right\rfloor,
		\end{aligned}
	\end{equation}
	where $\bar{\varphi}_\mathrm{max}$ is the maximum of the function $\bar{\varphi}(\bar{\varphi}_F)$ for $\bar{\varphi}_F \in [-\pi,\pi)$ and $\left\lfloor\cdot\right\rfloor$ represents the floor function. 
	As for each $n\in\left\{[N_\downarrow, N_\uparrow],\;n\neq0\right\}$ there will be two solutions, the total number of operating points, including the one for $n=0$, is given by
	\begin{equation}
		\label{eq:number_of_solutions_1}
		N_\mathrm{tot} = 1 + 2\left(N_\uparrow - N_\downarrow\right).
	\end{equation}
	Note that $N_\mathrm{tot}$ from the last expression is the total number of operating points, both stable (local minima) and unstable (local maxima). The number of stable operating points is given by
	\begin{equation}
		\label{eq:number_of_stable_solutions}
		\breve{N}_\mathrm{tot} = 1 + \left(N_\uparrow - N_\downarrow\right).
	\end{equation}
	In the next section, we will show how to apply the so far discussed technique to describe the low-energy properties of the Hamiltonian \eqref{eq:generic_hamiltonian}.
	
	\section{Series expansion of the effective Hamiltonian of a capacitively shunted superconducting dipole}
	Common experimental devices based on capacitively shunted superconducting dipoles \cite{manucharyan_fluxonium_2009, frattini_3-wave_2017, zhou_modular_2022, grimm_stabilization_2020, smith_superconducting_2020} are usually described by phenomenological, effective Hamiltonians which neglect the presence of the loop internal resonant modes. Under such approximation, the average values and the quantum fluctuations of the phases $\varphi_k$ describing the internal nodes of the dipole are considered purely dependent on the values and the quantum fluctuations of the phase $\varphi_0$ describing the main node of the dipole. This assumption is reasonable when the charging energy $E_{C_k}$ of the k-th JJ in the loop is much smaller than the charging energy $E_C$ of the external capacitance shunting the dipole, and it becomes exact in the limit $E_{C_k} \rightarrow 0$ \cite{rymarz_consistent_2022}. Note that, when an arm of the dipole includes a JJ in series with a large inductance, neglecting the internal JJ capacitances can result in a branched Hamiltonian which wouldn't correctly describe the dynamics of the circuit. In such case, a Born-Oppenheimer approximation can be applied to \eqref{eq:generic_hamiltonian}, accurately accounting for the presence of vanishingly small JJs' capacitances \cite{rymarz_consistent_2022}.
	In scenarios where the effect of the JJs' capacitances can be neglected, the Hamiltonian \eqref{eq:generic_hamiltonian} can be replaced by the one-body effective Hamiltonian
	\begin{equation}
		\label{eq:hamiltonian_eff}
		H_\mathrm{eff} = 4E_Cn_0^2 + U_\mathrm{eff}(\varphi_0,\bar{\varphi}_\mathrm{e}),
	\end{equation}
	where $E_C$ is the charging energy of the shunt capacitance shunting the dipole and $U_\mathrm{eff}$ is a phenomenological potential energy function which models the entire dipole as an effective, flux-tunable nonlinear inductor.
	A general expression for $U_\mathrm{eff}$ can be obtained from the potential energy $U_\mathrm{ind}$ in \eqref{eq:generic_hamiltonian} by imposing that the phase of the $k$-th internal node of the dipole $\varphi_k$ is related to the phase of the main node of the dipole $\varphi_0$ via a function $f_k$ such that $\varphi_k = f_k(\varphi_0)$, resulting in
	\begin{equation}
		\label{eq:potential_eff}
		U_\mathrm{eff}(\varphi_0,\bar{\varphi}_\mathrm{e}) = U_\mathrm{ind}([\varphi_0, f_1(\varphi_0),\dots,f_N(\varphi_0)],\bar{\varphi}_\mathrm{e}).
	\end{equation}
 	The function $f_k$ enforces the constraints imposed by Kirchhoff's laws to $\varphi_k$ in absence of the JJs' capacitances, and has an analytical form only for a narrow set of dipoles.
	If the functional form of $U_\mathrm{eff}$ is known, the effective Hamiltonian \eqref{eq:hamiltonian_eff} can be expanded, for a fixed value of external flux, around a flux-dependent equilibrium point $\bar{\varphi}_0$.
	The charge and phase variables $n_0$ and $\varphi_0$ can then be promoted to the operators $\hat{n}_0$ and $\hat{\varphi}_0$, respectively, yielding the series expansion of the effective Hamiltonian operator
	\begin{equation}
		\label{eq:hamiltonian_expanded}
		\hat{H}_\mathrm{eff} = 4E_C\hat{n}_0^2 + \frac{u_2}{2}\left(\hat{\varphi}_0-\bar{\varphi}_0\right)^2 + \sum_{n=3}^\infty\frac{u_n}{n!}\left(\hat{\varphi}_0-\bar{\varphi}_0\right)^n,
	\end{equation}
	where $u_n = \left.\frac{d^n U_\mathrm{eff}}{\varphi^n}\right|_{\varphi = \bar{\varphi}_0}$ are the Taylor expansion coefficients of $U_\mathrm{eff}$ around the equilibrium point $\bar{\varphi}_0$.
	Furthermore, it is possible to introduce the ladder operators $\hat{a}$ and $\hat{a}^\dagger$ satisfying the relations
	\begin{equation}
		\begin{aligned}
			\hat{\varphi}_0 - \bar{\varphi}_0 &= \varphi_\mathrm{zpf}(\hat{a} + \hat{a}^\dagger)\\
			\hat{n}_0 &= i n_\mathrm{zpf}(\hat{a} - \hat{a}^\dagger)\\
		\end{aligned}
	\end{equation}
	where 
	\begin{equation}
		\label{eq:zpf_quantities}
		\begin{aligned}
			\varphi_\mathrm{zpf}&=\frac{1}{\sqrt{2}}\left(\frac{8E_C}{u_2}\right)^\frac{1}{4}\\
			n_\mathrm{zpf}&=\frac{1}{\sqrt{2}}\left(\frac{u_2}{8E_C}\right)^\frac{1}{4}\\
		\end{aligned}
	\end{equation}
	are the quantum ground state uncertainties in phase and charge number, respectively.
	In the basis of these ladder operators, the Hamiltonian \eqref{eq:hamiltonian_expanded} reads
	\begin{equation}
		\label{eq:hamiltonian_expanded_ladder}
		\frac{\hat{H}_\mathrm{eff}}{\hbar} = \omega \hat{a}^\dagger \hat{a} + \sum_{n=3}^\infty g_n\left(\hat{a} + \hat{a}^\dagger\right)^n,
	\end{equation}
	where 
	\begin{equation}
		\label{eq:natural_frequency}
		\omega = \frac{\sqrt{8E_Cu_2}}{\hbar}
	\end{equation}
	is the natural frequency of the associated harmonic oscillator and
	\begin{equation}
		\label{eq:nonlinear_rates}
		g_n = \frac{\varphi_\mathrm{zpf}^n}{\hbar}\frac{u_n}{n!}
	\end{equation}
	are the rates of $n$-photon interactions.
	Expression \eqref{eq:hamiltonian_expanded_ladder} has been successfully applied to quantum circuits as the SNAIL \cite{frattini_3-wave_2017,grimm_stabilization_2020, ranadive_kerr_2022, zhou_modular_2022, chapman_high_2022}.
	However, in the case of an arbitrary superconducting dipole, the effective potential energy function $U_\mathrm{eff}(\varphi_0,\bar{\varphi}_\mathrm{e})$ might not have an analytical expression to begin with. 
	To overcome such obstacle, many aforementioned flux-biased loops were arranged in configurations which exploited certain symmetries and approximations to grant an analytical expression for $U_\mathrm{eff}$. 
	
	In this section, we show how to express \eqref{eq:hamiltonian_expanded_ladder} \textbf{without an a-priori knowledge} of the function $U_\mathrm{eff}(\varphi_0,\bar{\varphi}_\mathrm{e})$.
	\begin{figure}
		\centering
		\includegraphics[width = \columnwidth]{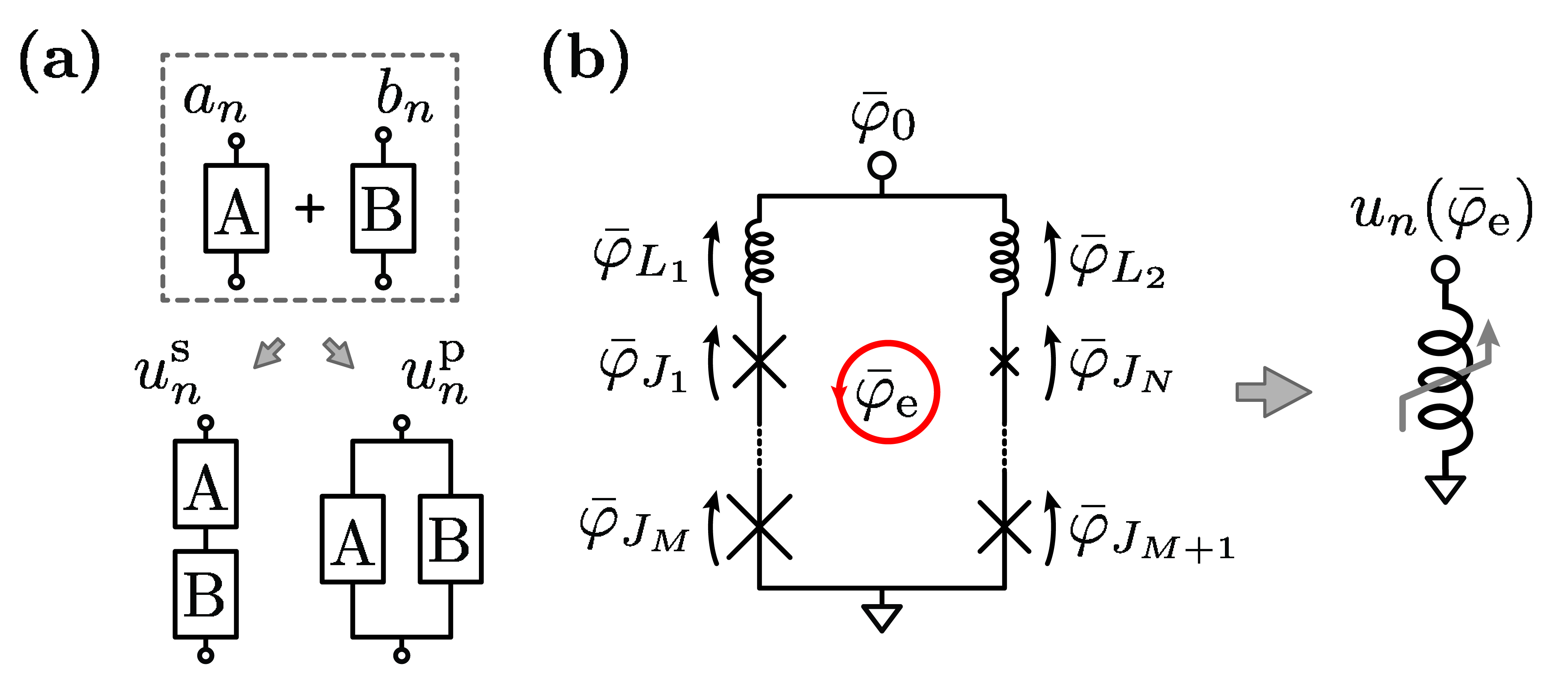}
		\caption{Reconstruction of the Taylor expansion coefficients for an arbitrary loop, neglecting the JJs' capacitances.
			(a)
			Two arbitrary nonlinear inductors \textbf{A} and \textbf{B} can be combined either in series or in parallel. Given their expansion coefficients, $a_n$ and $b_n$ respectively, it is possible to formulate the expansion coefficients resulting from their series combination $u_n^\mathrm{s}$, as well as those describing their parallel combination $u_n^\mathrm{p}$.
			(b)
			Neglecting the internal capacitances of a superconducting loop, the rules to combine nonlinear inductor expansion coefficients in series and in parallel can be applied to compute the expansion coefficients of the dipole without an a-priori knowledge of its potential energy function.
		}
		\label{fig:6}
	\end{figure}
	As described in section II, the equilibrium phase of each element in the loop is an analytical function of $\bar{\varphi}_F$. 
	Consequently, the $n$-th order potential energy expansion coefficient of the $x$-th element in the loop reads
	\begin{equation}
		x_n(\bar{\varphi}_F) = \left[\frac{d^n}{d\varphi^n}U_x(\varphi)\right]_{\varphi = \bar{\varphi}_{x}(\bar{\varphi}_F)}
	\end{equation}
	where $U_x(\varphi)$ is the potential energy function of the element $x$ and $ \bar{\varphi}_x(\bar{\varphi}_F)$ is the relation between the phase of the element $x$ of the loop and the phase of the free JJ.
	The individual expansion coefficients can be combined to return those of the effective potential energy $u_n$ in the following way.
	As a first step, it is useful to describe how the two arbitrary nonlinear inductors \textbf{A} and \textbf{B} in Fig. \ref{fig:6}(a) combine in series and in parallel, see details in appendix A. Given the expansion coefficients $a_n$ and $b_n$, their parallel combination is described by the expansion coefficients 
	\begin{equation}
		\label{eq:parallel_expansion}
		u^\mathrm{p}_n = a_n + b_n, 
	\end{equation}
	while the series combination can be expressed as 
	\begin{equation}
		\label{eq:series_expansion}
		u^\mathrm{s}_n = \mathrm{S}_n(\vec{a}_n, \vec{b}_n),
	\end{equation}
	where $\vec{a}_n = (a_0, a_1, \dots, a_n)$ and $\vec{b}_n = (a_0, a_1, \dots, a_n)$ are vectors whose components are the potential energy expansion coefficients of \textbf{A} and \textbf{B} up to order n, and $\mathrm{S}_n$ is a rational function of such components.
	The function $\mathrm{S}_n$ is derived in Appendix A and can be extended to an arbitrary number of elements, thus can be applied to both branches of the loop in Fig. \ref{fig:6} (b).
	Consequently, the expansion coefficients of the left and right branches of the dipole can be expressed as a function of $\bar{\varphi}_F$
	\begin{equation}
		\label{eq:branch_coefficients}
		\begin{aligned}
			u_n^\mathrm{left}(\bar{\varphi}_F) &=(-1)^n\mathrm{S}_n(\vec{x}_{n}(\bar{\varphi}_F), x\in\{{L_1,J_1,\dots,J_M}\})\\
			u_n^\mathrm{right}(\bar{\varphi}_F) &=\mathrm{S}_n(\vec{x}_{n}(\bar{\varphi}_F), x\in\{{L_{2},J_{M+1},\dots,J_N}\}).
		\end{aligned}
	\end{equation}
	The expansion coefficients of the dipole read
	\begin{equation}
		\label{eq:loop_coefficients}
		u_n(\bar{\varphi}_F) = u_n^\mathrm{left}(\bar{\varphi}_F) + u_n^\mathrm{right}(\bar{\varphi}_F).
	\end{equation}
	Notice how the relations \eqref{eq:phases_constrained} were obtained with the reference directions for phases and flux in Fig. \ref{fig:1}(d), while to assembly the dipole potential energy expansion coefficients $u_n$ it is more practical to use the reference directions in Fig. \ref{fig:6}(b).
	For a consistent computation of $u_n$, the functions $\bar{\varphi}_x(\bar{\varphi}_\mathrm{e})$ related to the left branch of the dipole in Fig. \ref{fig:1}(d) acquire a negative sign which only affects the odd expansion coefficients. This consideration gives rise to the $(-1)^n$ factor in the first line of \eqref{eq:branch_coefficients}.
	
	The relation between the expansion coefficients \eqref{eq:loop_coefficients} and external flux $\bar{\varphi}_\mathrm{e}$ can be represented as a parametric relation in $\bar{\varphi}_F$, similarly to the arbitrary branch CPR \eqref{eq:general_CPR}.
	Such description can also be applied to the parameters of Hamiltonian \eqref{eq:hamiltonian_expanded}, which are functions of the expansion coefficients $u_n$ according to expressions \eqref{eq:natural_frequency} and \eqref{eq:nonlinear_rates}.

	\section{Outlook}
	Under the one-body approximation for a generic flux-biased superconducting dipole, the method discussed in this manuscript makes it possible to express the effective Hamiltonian expansion coefficients \eqref{eq:natural_frequency} and \eqref{eq:nonlinear_rates} as well as the external flux in the first line of \eqref{eq:general_CPR} as analytical functions of the dipole design parameters and the free JJ equilibrium phase drop.
	This description will be at the core of a superconducting quantum circuits optimizer/synthesizer based on a gradient descend algorithm, which is currently under development. For instance, when designing a single-minima effective Hamiltonian to implement multi-photon parametric processes, one could specify constraints on the Hamiltonian parameters $\omega$ \eqref{eq:natural_frequency} and $g_n$ \eqref{eq:nonlinear_rates}. For a given dipole, the optimizer will be able to find the set of optimal parameters for the inductive elements within the loop, the optimal value of shunt capacitance $C_\mathrm{opt}$ and the optimal equilibrium phase for the free JJ $\bar{\varphi}_\mathrm{F_{opt}}$. From this values, the correspondent value of external flux $\bar{\varphi}_\mathrm{e}(\bar{\varphi}_\mathrm{F_{opt}})$ can be computed from the second equation in \eqref{eq:CPR_parametric}.
	
	In scenarios where the one-body approximation fails to describe the physical effects of interest of an arbitrary dipole, the equilibrium points computed with our method can be used as an input to more sophisticated algorithms.
	In particular, a recent proposal \cite{weiss_variational_2021} demonstrated how a variational tight-binding method can be applied to compute the properties of a large, multi-minima flux-biased superconducting circuit, accounting also for the junction capacitances. Once the expansion coefficients around each minima of the circuit are known, this approach provides a reduction in complexity with respect to a brute-force diagonalization of the Hamiltonian \eqref{eq:generic_hamiltonian}, with a promising fidelity in the estimation of the parameters of interest.
	A Born-Oppenheimer approximation can also be used to obtain an effective one-body Hamiltonian which accounts for the multi-body nature of an arbitrary dipole \cite{rymarz_consistent_2022}.
	
	Another natural application of our technique would be to model the repercussion of stray linear inductors and fabrication uncertainties on the Hamiltonian expansion coefficients of a superconducting dipole. Indeed, in the current approach the effective Hamiltonian \eqref{eq:hamiltonian_expanded} can hardly be computed in presence of stray inductors and asymmetry in arrays of JJs without recurring to the N+1 body Hamiltonian \eqref{eq:generic_hamiltonian}. Instead, our method provides a useful shortcut to correctly describe these scenarios.
	
	We are currently characterizing experimental devices with engineered asymmetries in arrays of JJs to validate the predictions of our method, with promising results. These will be presented in a future work focused on the Hamiltonian engineering capabilities enabled by the theory developed in this manuscript.

	\section{Conclusions}
	In this article, we have investigated the equilibrium properties of an arbitrary flux-biased superconducting loop. In particular, we demonstrated how the relation between the equilibrium points of a superconducting loop and its flux-bias can be expressed analytically as a parametric relation. We also derived a set of rules to count the number of local minima of the loop Hamiltonian. This approach can be generalized to circuits with more than one loop. As an immediate application of our technique, we showed how to compute the effective Hamiltonian Taylor expansion coefficients for an arbitrary flux-biased superconducting dipole shunted by an external capacitor. This enables quantitative analysis of yet-unexplored circuit topologies, overcoming the symmetry constraints and approximations which have limited the variety of devices investigated so far in literature. Our method is also suitable to implement a hardware-level Hamiltonian optimizer based on iterative algorithms: constraints on the quantities of interest, gradients and Hessians can all be expressed as analytical functions of the electrical parameters of the circuit.
	We believe that the method reported here could play an important role in the development of the next-generations of superconducting quantum devices, providing a shortcut towards the physical understanding, modeling and optimization of advanced flux-biased Josephson circuits.
	
	\section*{Acknowledgments}
	We thank D. P. DiVincenzo, J. Koch, G. Miano, Z. K. Minev, D. Weiss and A. B. Zorin for useful comments on the manuscript.
	This research was supported by the U.S. Army Research Office (ARO) under grant numbers W911NF-18-1-0212 and W911NF-16-1-0349, and by the National Science Foundation (NSF) under grant numbers 1941583 (ERC for CQN), and 2124511 (CCI for CQD-MQD). The views and conclusions contained in this document are those of the authors and should not be interpreted as representing official policies, either expressed or implied, of the grant agencies or the U.S. Government. The U.S. Government is authorized to reproduce and distribute reprints for Government purpose notwithstanding any copyright notation herein. 
	L.F. and M.H.D. are founders and L.F. is a shareholder of Quantum Circuits, Inc.
	
	\section*{Appendix A: Expansion coefficients of series and parallel configurations}
	The results from Section IV rely on the capability of computing the potential energy expansion coefficients of two arbitrary nonlinear inductors arranged in either parallel or series.
	Here, we show how to compute such coefficients as a function of those of the individual inductors.
	In general, for both configurations in Fig. \ref{fig:7}, the node phase $\varphi$, as well as the inductors phases $\varphi_\mathrm{A}$ and $\varphi_\mathrm{B}$ are related to their voltages, respectively $V$, $V_\mathrm{A}$ and $V_\mathrm{B}$ via the relations
	\begin{equation}
		\label{eq:branches_phase_voltage}
		\begin{aligned}
			\varphi(t) & = \frac{2\pi}{\Phi_0}\int_{-\infty}^t{V(t')dt'} + \bar{\varphi}\\
			\varphi_\mathrm{A}(t) & = \frac{2\pi}{\Phi_0} \int_{-\infty}^t{V_\mathrm{A}(t')dt'} + \bar{\varphi}_\mathrm{A}\\
			\varphi_\mathrm{B}(t) & = \frac{2\pi}{\Phi_0} \int_{-\infty}^t{V_\mathrm{B}(t')dt'} + \bar{\varphi}_\mathrm{B},\\
		\end{aligned}
	\end{equation}
	where $\bar{\varphi}$, $\bar{\varphi}_\mathrm{A}$ and $\bar{\varphi}_\mathrm{B}$ are, in our context, the equilibrium phases imposed by the external DC flux-bias to the node and to each inductor.
	To simplify the notation in the following treatment, we introduce the perturbations around the equilibrium for each phase, namely
	\begin{equation}
		\label{eq:general_phase_perturbations}
		\begin{aligned}
			\tilde{\varphi}_\mathrm{A} & = \varphi_\mathrm{A} - \bar{\varphi}_\mathrm{A}\\
			\tilde{\varphi}_\mathrm{B} & = \varphi_\mathrm{B} - \bar{\varphi}_\mathrm{B}\\
			\tilde{\varphi} & = \varphi - \bar{\varphi}.
		\end{aligned}
	\end{equation}
	Consequently, the expanded potential energy functions for the parallel and series configuration read
	\begin{equation}
		\label{eq:parallel_potential_expansion}
		\tilde{U}^\mathrm{p}(\tilde{\varphi}) = \sum_{n=1}^\infty \frac{u_n^\mathrm{p}}{n!}\tilde{\varphi}^n
	\end{equation}
	and
	\begin{equation}
		\label{eq:series_potential_expansion}
		\tilde{U}^\mathrm{s}(\tilde{\varphi}) = \sum_{n=1}^\infty \frac{u_n^\mathrm{s}}{n!}\tilde{\varphi}^n,
	\end{equation}
	while those of the nonlinear inductors \textbf{A} and \textbf{B} are
	\begin{equation}
		\label{eq:branches_potential_expansion}
		\begin{aligned}
			\tilde{U}_\mathrm{A}(\tilde{\varphi}_\mathrm{A}) & = \sum_{n=1}^\infty \frac{a_n}{n!}\tilde{\varphi}_\mathrm{A}^n\\
			\tilde{U}_\mathrm{B}(\tilde{\varphi}_\mathrm{B}) & = \sum_{n=1}^\infty \frac{b_n}{n!}\tilde{\varphi}_\mathrm{B}^n.
		\end{aligned}
	\end{equation}
	Note that the expanded potential energy functions are marked with a $\smallsim$ symbol to distinguish them from the non-expanded ones. The latter are used in Appendix B to characterize an arbitrary dipole in presence of a time-dependent external flux.

	\subsection*{Parallel expansion coefficients}
\begin{figure}
		\centering
		\includegraphics[width = \columnwidth]{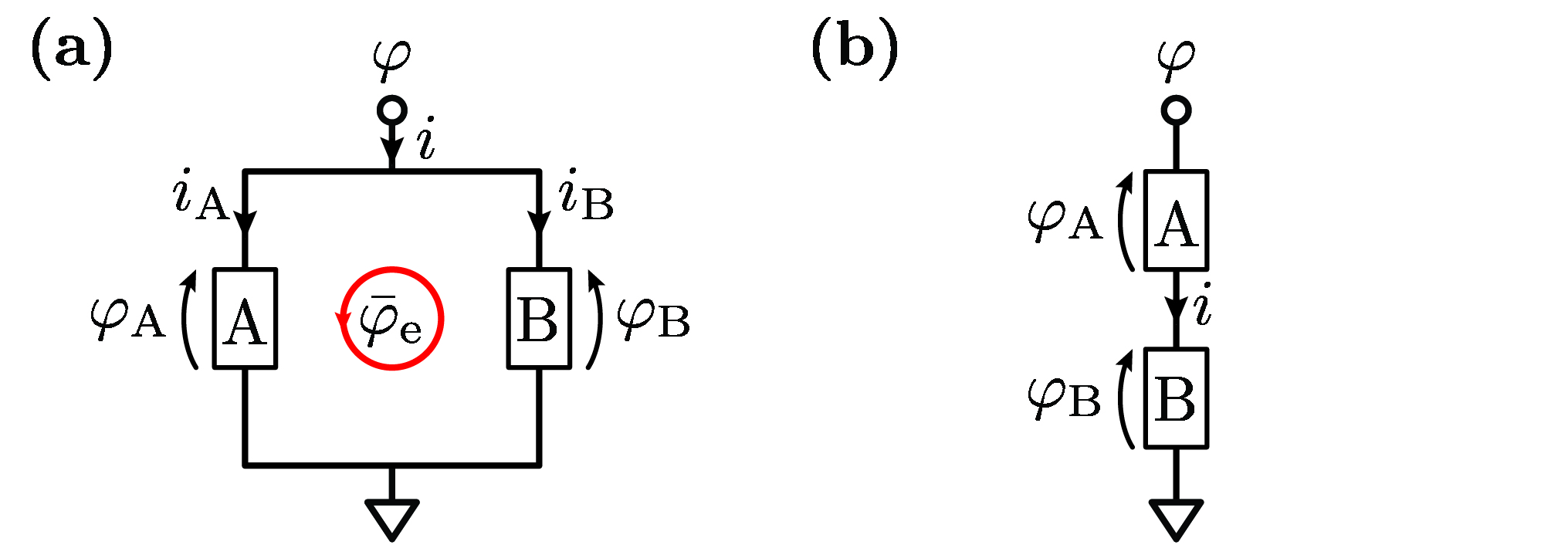}
		\caption{Two arbitrary nonlinear inductors \textbf{A} and \textbf{B} arranged in parallel and in series.
			(a) Parallel configuration of the two nonlinear inductors, forming a loop threaded with a external DC flux-bias $\bar{\varphi}_\mathrm{e}$. The phase drops across each inductor $\varphi_\mathrm{A}$ and $\varphi_\mathrm{B}$ are functions of both the loop node phase $\varphi$ and $\bar{\varphi}_\mathrm{e}$. The net current through the loop is the sum of the currents through each inductor, $i = i_\mathrm{A} + i_\mathrm{B}$.
			(b) Series configuration of the two nonlinear inductors. The phase drops across each inductor $\varphi_\mathrm{A}$ and $\varphi_\mathrm{B}$ are functions of the series node phase $\varphi$, with the constraint $\varphi = \varphi_\mathrm{A} + \varphi_\mathrm{B}$. The net current through the series is the same as the current flowing through each inductor.
		}
		\label{fig:7}
	\end{figure}

	To compute the expansion coefficients $u_n^\mathrm{p}$ in \eqref{eq:parallel_potential_expansion}, we notice how, with reference to Fig. \ref{fig:7}(a), \textbf{A} and \textbf{B} being in parallel implies $V_\mathrm{A} = V_\mathrm{B} = V$.
	Consequently, from Eq. \eqref{eq:branches_phase_voltage} and \eqref{eq:general_phase_perturbations} results that all the phase perturbations are the same in the parallel configuration
	\begin{equation}
		\label{eq:loop_phase_perturbations}
		\tilde{\varphi}_\mathrm{A} = \tilde{\varphi}_\mathrm{B} = \tilde{\varphi}.
	\end{equation}
	Thus, the parallel potential energy function reads
	\begin{equation}
		\label{eq:parallel_potential_energy}
		\tilde{U}^\mathrm{p}(\tilde{\varphi}) = \tilde{U}_\mathrm{A}(\tilde{\varphi}) + \tilde{U}_\mathrm{B}(\tilde{\varphi}).
	\end{equation}
	Computing the n-th order derivative of this last identity, and evaluating it for $\tilde{\varphi} = 0$, results in the expression \eqref{eq:parallel_expansion} from the definitions in \eqref{eq:parallel_potential_expansion} and \eqref{eq:branches_potential_expansion}.
	The generalization to a parallel configuration of an arbitrary set of nonlinear inductors $\{\textbf{A},\textbf{B},\textbf{C},\dots\}$ trivially reads
	\begin{equation}
		\label{eq:parallel_coefficients_generalized}
		u^\mathrm{p}_n = \sum\limits_{x\in\{a,b,c,\dots\}}x_n.
	\end{equation}
	\subsection*{Series expansion coefficients}
	The series configuration of \textbf{A} and \textbf{B} in Fig. \ref{fig:7}(b) is described by the set of constraints
	\begin{equation}
		\label{eq:series_constraints}
		\begin{aligned}
			\tilde{\varphi} = \tilde{\varphi}_\mathrm{A}(\tilde{\varphi}) & + \tilde{\varphi}_\mathrm{B}(\tilde{\varphi})\\
			\tilde{i}_\mathrm{A}[\tilde{\varphi}_\mathrm{A}(\tilde{\varphi})] & = \tilde{i}_\mathrm{B}[\tilde{\varphi}_\mathrm{B}(\tilde{\varphi})] 
		\end{aligned}
	\end{equation}
	where the first identity states the voltage conservation $V = V_\mathrm{A} + V_\mathrm{B}$, while the second identity imposes current conservation arising from the series arrangement.
	The phase perturbations of each inductor can be expanded as a function of the node phase as
	\begin{equation}
		\label{eq:participation_ratios}
		\begin{aligned}
			\tilde{\varphi}_\mathrm{A}(\tilde{\varphi}) = \sum_{n=1}^\infty \frac{p_{\mathrm{A}_n}}{n!}\tilde{\varphi}^{n}\\
			\tilde{\varphi}_\mathrm{B}(\tilde{\varphi}) = \sum_{n=1}^\infty \frac{p_{\mathrm{B}_n}}{n!}\tilde{\varphi}^{n}\\
		\end{aligned}
	\end{equation}
	where $p_{\mathrm{A}_n} = \left.\frac{d^n\tilde{\varphi}_\mathrm{A}}{d\tilde{\varphi}^n}\right|_{\tilde{\varphi} = 0}$ and $p_{\mathrm{B}_n} = \left.\frac{d^n\tilde{\varphi}_\mathrm{B}}{d\tilde{\varphi}^n}\right|_{\tilde{\varphi} = 0}$ are the nonlinear participation ratios \cite{frattini_optimizing_2018} of \textbf{A} and \textbf{B}.
	We now define the algebraic rules for the expansion coefficients
	\begin{equation}
		\label{eq:algebraic_rules}
		\begin{aligned}
			\frac{dp_{\mathrm{A}_n}}{d\tilde{\varphi}} & \eqdef \left.\frac{d}{d\tilde{\varphi}}\left(\frac{d^n\tilde{\varphi}_\mathrm{A}}{d\tilde{\varphi}^n}\right)\right|_{\tilde{\varphi} = 0} = p_{\mathrm{A}_{n+1}}\\
			\frac{dp_{\mathrm{B}_n}}{d\tilde{\varphi}} & \eqdef \left.\frac{d}{d\tilde{\varphi}}\left(\frac{d^n\tilde{\varphi}_\mathrm{B}}{d\tilde{\varphi}^n}\right)\right|_{\tilde{\varphi} = 0} = p_{\mathrm{B}_{n+1}}\\
			\frac{du^\mathrm{p}_n}{d\tilde{\varphi}} & \eqdef \left.\frac{d}{d\tilde{\varphi}}\left(\frac{d^n\tilde{U}^\mathrm{p}}{d\tilde{\varphi}^n}\right)\right|_{\tilde{\varphi} = 0} = u^\mathrm{p}_{n+1}\\
			\frac{du^\mathrm{s}_n}{d\tilde{\varphi}} & \eqdef \left.\frac{d}{d\tilde{\varphi}}\left(\frac{d^n\tilde{U}^\mathrm{s}}{d\tilde{\varphi}^n}\right)\right|_{\tilde{\varphi} = 0} = u^\mathrm{s}_{n+1}\\
			\frac{da_n}{d\tilde{\varphi}} & \eqdef \left.\frac{d}{d\tilde{\varphi}}\left(\frac{d^n\tilde{U}_\mathrm{A}}{d\tilde{\varphi}_\mathrm{A}^n}\right)\right|_{\tilde{\varphi} = 0} = p_{\mathrm{A}_1}a_{n+1}\\
			\frac{da_n}{d\tilde{\varphi}} & \eqdef \left.\frac{d}{d\tilde{\varphi}}\left(\frac{d^n\tilde{U}_\mathrm{B}}{d\tilde{\varphi}_\mathrm{B}^n}\right)\right|_{\tilde{\varphi} = 0} = p_{\mathrm{B}_1}b_{n+1}\\
		\end{aligned}
	\end{equation}
	where we implicitly mean that the derivative is taken \textbf{before} evaluating the corresponding functions for $\tilde{\varphi} = 0$.
	This abuse of notation will ease the elaboration of the $S_n$ function.
	By computing the n-th order derivatives with respect to $\tilde{\varphi}$ of the first line of \eqref{eq:series_constraints}, and combining them with the expressions \eqref{eq:participation_ratios}, we obtain the constraints on the participation ratios
	\begin{equation}
		\label{eq:participation_ratios_constrained}
		\begin{aligned}
			p_{\mathrm{A}_1} + p_{\mathrm{B}_1} & = 1\\
			p_{\mathrm{A}_n} + p_{\mathrm{B}_n} & = 0 \;\;\; n\geq 2.
		\end{aligned}
	\end{equation}
	The currents through each nonlinear inductor are related to their potential energy functions via the relations
	\begin{equation}
		\label{eq:currents_potentials}
		\begin{aligned}
			\tilde{i}_\mathrm{A}(\tilde{\varphi}_\mathrm{A}) = \frac{2\pi}{\Phi_0}\frac{d\tilde{U}_\mathrm{A}}{d\tilde{\varphi}_\mathrm{A}}\\
			\tilde{i}_\mathrm{B}(\tilde{\varphi}_\mathrm{B}) = \frac{2\pi}{\Phi_0}\frac{d\tilde{U}_\mathrm{B}}{d\tilde{\varphi}_\mathrm{B}}.
		\end{aligned}
	\end{equation}
	Combining these last expressions with the constraints in the second line of \eqref{eq:series_constraints}, and computing the derivative with respect to $\tilde{\varphi}$, we obtain
	\begin{equation}
		\label{eq:current_conservation_expanded}
		p_{\mathrm{A}_1}a_2 = p_{\mathrm{B}_1}b_2
	\end{equation}
	which, together with the first of \eqref{eq:participation_ratios_constrained}, provides the closed expression for the linear participation ratios
	\begin{equation}
		\label{eq:participation_ratios_linear}
		\begin{aligned}
			p_{\mathrm{A}_1} & = \frac{b_2}{a_2 + b_2}\\
			p_{\mathrm{B}_1} & = \frac{a_2}{a_2 + b_2}.
		\end{aligned}
	\end{equation}
	Higher order participation ratios can be retrieved by applying the algebraic rules \eqref{eq:algebraic_rules} to the identity \eqref{eq:current_conservation_expanded}. Their expressions up to third order are
	\begin{equation}
		\label{eq:participation_ratios_higher_order}
		\begin{aligned}
			p_{\mathrm{A}_2} & = \frac{p_{\mathrm{B}_1}^2b_3 - p_{\mathrm{A}_1}^2a_3}{a_2 + b_2}\\
			p_{\mathrm{B}_2} & = \frac{p_{\mathrm{A}_1}^2a_3 - p_{\mathrm{B}_1}^2b_3}{a_2 + b_2}\\
			p_{\mathrm{A}_3} & = \frac{p_{\mathrm{B}_1}^3b_4 - p_{\mathrm{A}_1}^3a_4 + 2(p_{\mathrm{B}_1}p_{\mathrm{B}_2}b_3 - p_{\mathrm{A}_1}p_{\mathrm{A}_2}a_3)}{(a_2 + b_2)^2}\\
			p_{\mathrm{B}_3} & = \frac{p_{\mathrm{A}_1}^3a_4 - p_{\mathrm{B}_1}^3b_4 + 2(p_{\mathrm{A}_1}p_{\mathrm{A}_2}a_3 - p_{\mathrm{B}_1}p_{\mathrm{B}_2}b_3)}{(a_2 + b_2)^2}
		\end{aligned}
	\end{equation}
	We can now retrieve the series potential energy expansion coefficients $u^\mathrm{s}_n$ as a function of $a_n$ and $b_n$.
	The series potential energy function can be expressed as 
	\begin{equation}
		\tilde{U}^\mathrm{s}(\tilde{\varphi}) = \tilde{U}_\mathrm{A}\left[\tilde{\varphi}_\mathrm{A}\left(\tilde{\varphi}\right)\right] + \tilde{U}_\mathrm{B}\left[\tilde{\varphi}_\mathrm{B}\left(\tilde{\varphi}\right)\right]
	\end{equation}
	from which the expression of $u^\mathrm{s}_1$ can be retrieved as
	\begin{equation}
		\label{eq:c1}
		u^\mathrm{s}_1 = p_{\mathrm{A}_1}a_1 + p_{\mathrm{B}_1}b_1.
	\end{equation}
	Applying the algebraic rules \eqref{eq:algebraic_rules} recursively to this last identity, and keeping in mind the constraints \eqref{eq:participation_ratios_constrained}, $u^\mathrm{s}_n$ can be easily computed. Here we show such coefficients up to fifth order
	\begin{equation}
		\label{eq:SN}
		\begin{aligned}
			u^\mathrm{s}_2 & = p_{\mathrm{A}_1}^2a_2 + p_{\mathrm{B}_1}^2b_2 \\
			u^\mathrm{s}_3 & = p_{\mathrm{A}_1}^3a_3 + p_{\mathrm{B}_1}^3b_3 \\
			u^\mathrm{s}_4 & = p_{\mathrm{A}_1}^4a_4 + p_{\mathrm{B}_1}^4b_4 + 3(p_{\mathrm{A}_1}^2p_{\mathrm{A}_2}a_3 + p_{\mathrm{B}_1}^2p_{\mathrm{B}_2}b_3)\\
			u^\mathrm{s}_5 & = p_{\mathrm{A}_1}^5a_5 + p_{\mathrm{B}_1}^5b_5 + 4(p_{\mathrm{A}_1}^3p_{\mathrm{A}_2}a_4 + p_{\mathrm{B}_1}^3p_{\mathrm{B}_2}b_4)\\
			&+3[(p_{\mathrm{A}_1}^2p_{\mathrm{A}_3} + 2p_{\mathrm{A}_1}p_{\mathrm{A}_2}^2)a_3 + p_{\mathrm{A}_1}^2p_{\mathrm{A}_3}a_4]\\
			&+3[(p_{\mathrm{B}_1}^2p_{\mathrm{B}_3} + 2p_{\mathrm{B}_1}p_{\mathrm{B}_2}^2)b_3 + p_{\mathrm{B}_1}^2p_{\mathrm{B}_3}b_4].
		\end{aligned}
	\end{equation}
	Notice how these expressions are invariant under the swap of \textbf{A} and \textbf{B}, as the properties of their series are independent from the order of arrangement. Consequently, they can be computed, for instance, just for \textbf{A} and then trivially extended to include \textbf{B} as well.
	
	The rational functions in \eqref{eq:SN} define the series combination $S_n$ in \eqref{eq:series_expansion}, and can be extended to an arbitrary set of series nonlinear inductors $\{\textbf{A},\textbf{B},\textbf{C},\dots\}$. Said $x_n$ the n-th order expansion coefficient of the $x$-th nonlinear inductor in the series, its linear participation ratio $p_{x_1}$ can be expressed as
	\begin{equation}
		\label{eq:generalized_participation}
		p_{x_1} = \frac{\prod\limits_{l\neq x}l_2}{\sum\limits_{l}\prod\limits_{m\neq l}m_2} 
	\end{equation}
	where $l,m\in\{a,b,c,\dots\}$.
	Higher order participation ratios for the $x$-th element can be computed from \eqref{eq:generalized_participation} by applying the algebraic rules \eqref{eq:algebraic_rules}.
	As an example, the second-order participation ratio of the $x$-th element in an arbitrary array reads
	\begin{equation}
		\label{eq:generalized_participation_2}
		\begin{aligned}
			p_{x_2} & = \frac{\left(\frac{d}{d\tilde{\varphi}}\prod\limits_{l\neq x}l_2\right)\left(\sum\limits_{l}\prod\limits_{m\neq l}m_2\right)}{\left(\sum\limits_{l}\prod\limits_{m\neq l}m_2\right)^2} \\
			& - \frac{\left(\prod\limits_{l\neq x}l_2\right)\left(\sum\limits_{l}\frac{d}{d\tilde{\varphi}}\prod\limits_{m\neq l}m_2\right)}{\left(\sum\limits_{l}\prod\limits_{m\neq l}m_2\right)^2}
		\end{aligned}
	\end{equation}
	where 
	\begin{equation}
		\frac{d}{d\tilde{\varphi}}\prod\limits_{x}x_2 = \left(\prod\limits_{x}x_2\right)\left(\sum\limits_{x}\frac{p_{x_1}x_3}{x_2}\right).
	\end{equation}
	For an arbitrary array, the expansion coefficients \eqref{eq:SN} up to fifth order generalize as
	\begin{equation}
		\label{eq:SN_generalized}
		\begin{aligned}
			u^\mathrm{s}_2 & = \sum_x p_{x_1}^2x_2\\
			u^\mathrm{s}_3 & = \sum_x p_{x_1}^3x_3\\
			u^\mathrm{s}_4 & = \sum_x p_{x_1}^4x_4 + 3p_{x_1}^2p_{x_2}x_3\\
			u^\mathrm{s}_5 & = \sum_x p_{x_1}^5x_5 + 4(p_{x_1}^3p_{x_2}x_4)\\
			&+3[(p_{x_1}^2p_{x_3} + 2p_{x_1}p_{x_2}^2)x_3 + p_{x_1}^2p_{x_3}x_4]
		\end{aligned}
	\end{equation}
	which, together with $p_{x_n}$, define the rational function $S_n$ for an arbitrary array in \eqref{eq:branch_coefficients}.
	When the expressions \eqref{eq:SN_generalized} are applied to the arm of a loop which doesn't contain the free JJ, they don't present any singularity as a function of $\bar\varphi_F$.
	This is easily demonstrated by noticing that $x_2(\bar{\varphi}_F)>0$ for any constrained element, and the denominators of $u^\mathrm{s}_n$ are multinomial functions of $x_2$ with positive coefficients.
	On the other hand, when a branch contains a free JJ, $u^\mathrm{s}_n$ can have singularities if the branch is multi-valued.
	
	\section*{Appendix B: Expansion coefficients with time-dependent external flux}
	In this section, we analyze the properties of a capacitively shunted superconducting dipole in presence of a time-dependent external magnetic flux. Thus, we introduce the time-dependent component of the external magnetic flux $\tilde{\varphi}_\mathrm{e}(t)$, such that the total flux $\varphi_\mathrm{e}(t)$ threaded to the loop forming the dipole reads
	\begin{equation}
	\label{eq:total_flux}
	\varphi_\mathrm{e}(t) = \bar{\varphi}_\mathrm{e} + \tilde{\varphi}_\mathrm{e}(t).	
	\end{equation}
	We will consider the case where $\max\left\{\tilde{\varphi}_\mathrm{e}(t)\right\}\ll2\pi$, thus the time-dependent component of the external magnetic flux can be treated perturbatively.	
	Introducing the maximum frequency of $\tilde{\varphi}_\mathrm{e}$ as $\omega^\mathrm{max}_{\tilde{\varphi}_\mathrm{e}}$, the analysis can be performed with two different approaches. 
	In the case where $\omega^\mathrm{max}_{\tilde{\varphi}_\mathrm{e}}\ll\omega$, where $\omega$ is the natural frequency of the harmonic oscillator associated to the capacitively shunted dipole as in Eq. \eqref{eq:natural_frequency},
	it is reasonable to assume that the effective potential energy \eqref{eq:potential_eff} acquires a slow time dependence, corresponding to an adiabatic modulation of its minima. This is a realistic assumption, for instance, to model the effects of low-frequency flux noise on the device.
	In the case where $\omega^\mathrm{max}_{\tilde{\varphi}_\mathrm{e}}\approx\omega$, the external magnetic flux can strongly modify the dynamics of the system, for instance, by driving parametric processes. As a consequence, the effective potential energy \eqref{eq:potential_eff} acquires an additional functional dependence on $\tilde{\varphi}_\mathrm{e}(t)$. We will refer to this scenario as ``flux pumping''.
	In the following treatment, we will refer to the circuit in Fig. \ref{fig:7}(a).
	\subsection*{Sensitivity to low-frequency flux noise}
	In realistic experimental scenarios involving DC flux-biased superconducting loops is important to account for the presence of a non-deterministic component of the total magnetic flux. Such component can be modeled as a stochastic signal with a low-frequency power spectral density, typically $1/f$ noise \cite{kumar_origin_2016}, which can arise from both magnetic impurities in the vicinity of the loops and room temperature electronics providing the DC flux-bias to the device.
	A major detrimental effect of such noise is a slow, random modulation of the effective Hamiltonian parameters in Eq.\eqref{eq:hamiltonian_expanded_ladder}, which can  cause dephasing events in flux-biased superconducting quantum devices. The first-order sensitivity of the Hamiltonian parameters to low-frequency flux noise can be evaluated by computing their derivatives with respect to the external DC flux $\bar{\varphi}_\mathrm{e}$. Here, we show how to compute such derivatives and relate them to the potential energy expansion coefficients of the nonlinear inductors \textbf{A} and \textbf{B} in Fig. \ref{fig:7}(a).
	The sensitivity to slow-flux modulations of the natural frequency in Eq. \eqref{eq:natural_frequency} can be expressed as
	\begin{equation}
		\label{eq:natural_frequency_flux_sensitivity}
		\frac{d\omega}{d\bar{\varphi}_\mathrm{e}} = \frac{1}{\hbar}\sqrt{\frac{4E_C}{u_2}}\frac{du_2}{d\bar{\varphi}_\mathrm{e}},
	\end{equation}
	while that of the n-photons interaction rates in Eq. \eqref{eq:nonlinear_rates} reads
		\begin{equation}
		\label{eq:nonlinear_rates_flux_sensitivity_partial}
		\frac{dg_n}{d\bar{\varphi}_\mathrm{e}} = \frac{1}{\hbar}\left[\frac{\varphi_\mathrm{ZPF}^{n-1}}{(n-1)!}\frac{d\varphi_\mathrm{ZPF}}{d\bar{\varphi}_\mathrm{e}}u_n + \varphi_\mathrm{ZPF}^n\frac{du_n}{d\bar{\varphi}_\mathrm{e}}\right],
	\end{equation}
	where
	\begin{equation}
		\label{eq:phiZPF_flux_sensitivity}
		\frac{d\varphi_\mathrm{ZPF}}{d\bar{\varphi}_\mathrm{e}} = -\frac{\varphi_\mathrm{ZPF}}{4u_2}\frac{du_2}{d\bar{\varphi}_\mathrm{e}}.
	\end{equation}
	Inserting this last expression in Eq. \eqref{eq:nonlinear_rates_flux_sensitivity_partial} results in
	\begin{equation}
	\label{eq:nonlinear_rates_flux_sensitivity}
		\frac{dg_n}{d\bar{\varphi}_\mathrm{e}} = 
		\frac{\varphi_\mathrm{ZPF}^n}{\hbar}\left[\frac{du_n}{d\bar{\varphi}_\mathrm{e}}
		-\frac{1}{4(n-1)!}\frac{u_n}{u_2}\frac{du_2}{d\bar{\varphi}_\mathrm{e}}\right].
	\end{equation}
	The derivative of the potential energy expansion coefficients with respect to $\bar{\varphi}_\mathrm{e}$ can be expressed as
	\begin{equation}
		\label{eq:un_flux_sensitivity_initial}
		\frac{du_n}{d\bar{\varphi}_\mathrm{e}} = 
		\frac{da_n}{d\bar{\varphi}_\mathrm{e}} + \frac{db_n}{d\bar{\varphi}_\mathrm{e}}.
	\end{equation}
	By noticing that
	\begin{equation}
		\label{eq:branches_expansion_coefficients}
		\begin{aligned}
			a_n & = \frac{d^nU_\mathrm{A}}{d\varphi_\mathrm{A}^n}[\bar{\varphi}_\mathrm{A}(\bar{\varphi}_\mathrm{e})]\\
			b_n & = \frac{d^nU_\mathrm{B}}{d\varphi_\mathrm{B}^n}[\bar{\varphi}_\mathrm{B}(\bar{\varphi}_\mathrm{e})],
		\end{aligned}
	\end{equation}
	where $U_A(\varphi_\mathrm{A})$ and $U_B(\varphi_\mathrm{B})$ are the potential energy functions of the left and right branches of the dipole, respectively, we derive the algebraic rules for computing the derivatives of the expansion coefficients with respect to the external DC flux-bias
	\begin{equation}
		\label{eq:flux_sensitivity_algebraic_rules}
		\begin{aligned}
			\frac{da_n}{d\bar{\varphi}_\mathrm{e}} & = \frac{d\bar{\varphi}_\mathrm{A}}{d\bar{\varphi}_\mathrm{e}}a_{n+1}\\
			\frac{db_n}{d\bar{\varphi}_\mathrm{e}} & = \frac{d\bar{\varphi}_\mathrm{B}}{d\bar{\varphi}_\mathrm{e}}b_{n+1}.
		\end{aligned}
	\end{equation}
	Inserting these last expressions in Eq. \eqref{eq:un_flux_sensitivity_initial} results in
	\begin{equation}
		\label{eq:un_flux_sensitivity_intermediate}
		\frac{du_n}{d\bar{\varphi}_\mathrm{e}} = 
		\frac{d\bar{\varphi}_\mathrm{A}}{d\bar{\varphi}_\mathrm{e}}a_{n+1} +
		\frac{d\bar{\varphi}_\mathrm{B}}{d\bar{\varphi}_\mathrm{e}}b_{n+1},
	\end{equation}
	which can be further processed to explicit the dependence on the free JJ equilibrium phase $\bar{\varphi}_\mathrm{F}$ as
		\begin{equation}
		\label{eq:un_flux_sensitivity}
		\frac{du_n}{d\bar{\varphi}_\mathrm{e}} = \left(\frac{d\bar{\varphi}_\mathrm{e}}{d\bar{\varphi}_\mathrm{F}}\right)^{-1}\left(
		\frac{d\bar{\varphi}_\mathrm{A}}{d\bar{\varphi}_\mathrm{F}}a_{n+1} +
		\frac{d\bar{\varphi}_\mathrm{B}}{d\bar{\varphi}_\mathrm{F}}b_{n+1}\right).
	\end{equation}
	Note how this last expression was obtained with the reference directions for phases as in Fig. \ref{fig:7}(a). As a consequence, the function $\bar{\varphi}_\mathrm{A}(\bar{\varphi}_\mathrm{F})$ has opposite sign with respect to the one that can be obtained following the method for computing the equilibrium phase drops in Section II, which relies on the reference directions for phases as in Fig.\ref{fig:2}(a).
	The first-order sensitivity to slow-flux modulations of the dipole' expansion coefficients in Eq. \eqref{eq:un_flux_sensitivity} can be inserted in the expressions \eqref{eq:natural_frequency_flux_sensitivity} and \eqref{eq:nonlinear_rates_flux_sensitivity} to retrieve analytical expressions for $\frac{d\omega}{d\bar{\varphi}_\mathrm{e}}$ and $\frac{dg_n}{d\bar{\varphi}_\mathrm{e}}$. Note how the latter are analytical functions of the free JJ equilibrium phase drop and the electrical parameters of the circuit, thus their relation to the external DC flux-bias $\bar{\varphi}_\mathrm{e}$ still has the form of a parametric curve of curvilinear parameter $\bar{\varphi}_\mathrm{F}$.
	The n-th order sensitivity to slow-flux modulations can be obtained by computing the n-th order derivative with respect to $\bar{\varphi}_\mathrm{e}$ of expressions \eqref{eq:natural_frequency_flux_sensitivity}, \eqref{eq:nonlinear_rates_flux_sensitivity} and \eqref{eq:un_flux_sensitivity} and applying the algebraic rules in \eqref{eq:flux_sensitivity_algebraic_rules}.
	\subsection*{Flux pumping}
	In this section, we analyze the case of a fast time-dependent flux drive $\tilde{\varphi}_\mathrm{e}(t)$ applied to the nonlinear dipole in Fig. \ref{fig:7}(a). The potential energy of such driven dipole can be expanded in Taylor series up to order $\mathcal{O}$ for both the main node phase $\tilde{\varphi}$ and the flux drive $\tilde{\varphi}_\mathrm{e}$
	\begin{equation}
		\label{eq:driven_potential_energy_expansion}
		\tilde{U}(\tilde{\varphi}, \tilde{\varphi}_\mathrm{e}) = \sum_{n=0}^{\mathcal{O}}\sum_{l=0}^{\mathcal{O}-n}\frac{u_{nl}}{n!l!}\tilde{\varphi}^n\tilde{\varphi}_\mathrm{e}^l
	\end{equation}
	where
	\begin{equation}
		\label{eq:driven_expansion_coefficients_intermediate}
	u_{nl} = \left.\frac{\partial^{n+l}U(\tilde{\varphi},\tilde{\varphi}_\mathrm{e},\bar{\varphi}_\mathrm{e})}{\partial\tilde{\varphi}^n\partial\tilde{\varphi}_\mathrm{e}^l}\right|_{(\tilde{\varphi},\tilde{\varphi}_\mathrm{e})=(0,0)}
	\end{equation}
	are the expansion coefficients of order $n+l$ of the driven dipole potential energy function $U(\tilde{\varphi},\tilde{\varphi}_\mathrm{e},\bar{\varphi}_\mathrm{e})$.
	Following the results in \cite{you_circuit_2019, riwar_circuit_2022, bryon_time-dependent_2023}, the flux-drive $\tilde{\varphi}_\mathrm{e}$ has to be allocated among the nonlinear inductors \textbf{A} and \textbf{B} according to the spatial distribution of the magnetic vector potential on the device. In particular, the phase drops $\varphi_\mathrm{A}$ and $\varphi_\mathrm{B}$ across the nonlinear inductors in Eq. \eqref{eq:branches_phase_voltage} acquire a functional dependence on $\tilde{\varphi}_\mathrm{e}$ of the form
	\begin{equation}
		\label{eq:phase_drops_flux_allocation}
		\begin{aligned}
			\varphi_\mathrm{A} & = \tilde{\varphi} - \alpha\tilde{\varphi}_\mathrm{e} + \bar{\varphi}_\mathrm{A}\\
			\varphi_\mathrm{B} & = \tilde{\varphi} + (1-\alpha)\tilde{\varphi}_\mathrm{e} + \bar{\varphi}_\mathrm{B}
		\end{aligned}
	\end{equation}
	where $\alpha$ is the allocation factor that can be computed following the procedure detailed in \cite{riwar_circuit_2022}.
	Note how, according to these last expressions, the difference between the phases across \textbf{B} and \textbf{A} reads
	\begin{equation}
		\varphi_\mathrm{B} - \varphi_\mathrm{A} = \bar{\varphi}_\mathrm{B} - \bar{\varphi}_\mathrm{A}  + \tilde{\varphi}_\mathrm{e} = \bar{\varphi}_\mathrm{e} + \tilde{\varphi}_\mathrm{e},
	\end{equation}
	resulting in the expression \eqref{eq:total_flux} for the total magnetic flux threaded to the loop. The last identity can be demonstrated by expressing the second equation in \eqref{eq:single_loop_KL} with the reference directions for the phase drops in Fig. \ref{fig:7}(a).
	With the expressions in \eqref{eq:phase_drops_flux_allocation}, the driven dipole potential energy function can be expressed as
	\begin{equation}
		\label{eq:driven_potential_energy}
		\begin{aligned}
			U(\tilde{\varphi},\tilde{\varphi}_\mathrm{e},\bar{\varphi}_\mathrm{e}) & = 
			U_\mathrm{A}(\tilde{\varphi} - \alpha\tilde{\varphi}_\mathrm{e} + \bar{\varphi}_\mathrm{A}) \\
			& + U_\mathrm{B}[\tilde{\varphi} + (1-\alpha)\tilde{\varphi}_\mathrm{e} + \bar{\varphi}_\mathrm{B}],
			\end{aligned}
	\end{equation}
	thus the dipole expansion coefficients in \eqref{eq:driven_expansion_coefficients_intermediate} read
	\begin{equation}
		\label{eq:driven_expansion_coefficients_parallel}
			u_{nl} = a_{nl} + b_{nl},
	\end{equation}
	where
	\begin{equation}
		\label{eq:driven_expansion_coefficients_branches_intermediate}
		\begin{aligned}
			a_{nl} & = \frac{\partial^{n+l}U_\mathrm{A}}{\partial\tilde{\varphi}^n\partial\tilde{\varphi}_\mathrm{e}^l}(\bar{\varphi}_\mathrm{A})\\
			b_{nl} & = \frac{\partial^{n+l}U_\mathrm{B}}{\partial\tilde{\varphi}^n\partial\tilde{\varphi}_\mathrm{e}^l}(\bar{\varphi}_\mathrm{B})
		\end{aligned}
	\end{equation}
	are the driven expansion coefficients of the nonlinear inductors \textbf{A} and \textbf{B}, respectively. From the functional dependence of $U_\mathrm{A}$ and $U_\mathrm{B}$ on the right side of expression \eqref{eq:driven_potential_energy}, the driven expansion coefficients in \eqref{eq:driven_expansion_coefficients_branches_intermediate} can be expressed as a function of the bare ones in Eq. \eqref{eq:branches_expansion_coefficients} as
	\begin{equation}
		\label{eq:driven_expansion_coefficients_branches}
		\begin{aligned}
		a_{nl} & = (-\alpha)^la_{n+l}\\
		b_{nl} & = (1-\alpha)^lb_{n+l},
		\end{aligned}
	\end{equation}
	resulting in a compact final expression for the driven dipole expansion coefficients \eqref{eq:driven_expansion_coefficients_intermediate}, which reads
	\begin{equation}
		\label{eq:driven_expansion_coefficients}
		u_{nl} = (-\alpha)^la_{n+l} + (1-\alpha)^lb_{n+l}.
	\end{equation}
	This last expression can be inserted in the driven dipole potential energy Taylor expansion \eqref{eq:driven_potential_energy_expansion} to obtain an expression as a function of the bare expansion coefficients of the nonlinear inductors \textbf{A} and \textbf{B}.
	The sensitivity to low-frequency flux noise of the driven dipole potential energy expansion coefficients in \eqref{eq:driven_expansion_coefficients} can be computed following the same treatment explained in the previous section as
		\begin{equation}
		\label{eq:driven_expansion_coefficients_flux_sensitivity}
		\begin{aligned}
		\frac{du_{nl}}{d\bar{\varphi}_\mathrm{e}} = \left(\frac{d\bar{\varphi}_\mathrm{e}}{d\bar{\varphi}_\mathrm{F}}\right)^{-1}
		&\left[(-\alpha)^l\frac{d\bar{\varphi}_\mathrm{A}}{d\bar{\varphi}_\mathrm{F}}a_{n+l+1}\right.\\
		&\left.+(1-\alpha)^l\frac{d\bar{\varphi}_\mathrm{B}}{d\bar{\varphi}_\mathrm{F}}b_{n+l+1}\right].
		\end{aligned}
	\end{equation}
	\bibliography{bibliography}
\end{document}